\documentclass[journal=jpcafh,manuscript=article,layout=onecolumn]{achemso}
\setkeys{acs}{articletitle = true}
\setkeys{acs}{chaptertitle = true}
\setkeys{acs}{abbreviations = false}
\setkeys{acs}{etalmode = truncate}
\setkeys{acs}{maxauthors = 10}
\SectionNumbersOn
\usepackage[version=3]{mhchem} % Formula subscripts using \ce{}

\usepackage{graphicx}% Include figure files
\usepackage{dcolumn}% Align table columns on decimal point
\usepackage{bm}% bold math
\usepackage[utf8]{inputenc}
\usepackage[T1]{fontenc}
\usepackage{mathptmx}
\usepackage{etoolbox}
\usepackage{longtable}
\usepackage[section]{placeins}

\usepackage{multirow}
\usepackage[caption=false]{subfig}

\usepackage{algorithm}
\usepackage{algpseudocode}
\usepackage{xr}
\title{New Implementation of 
an Equation-of-Motion
Coupled-Cluster Damped-Response 
Framework with Illustrative Applications 
to Resonant Inelastic X-ray Scattering
}% Force line breaks with \\
\author{Anna Kristina Schnack-Petersen}
 \email{akrsc@kemi.dtu.dk}
 \affiliation[DTU]
 {DTU Chemistry, Technical University of Denmark, DK-2800 Kongens Lyngby, Denmark}
 \author{Torsha Moitra}
 \affiliation[DTU]
 {DTU Chemistry, Technical University of Denmark, DK-2800 Kongens Lyngby, Denmark}
  \alsoaffiliation[UiT]{Hylleraas Centre for Quantum Molecular Sciences, Department of Chemistry, UiT – The Arctic University of Norway, 9037 Troms{\o}, Norway}
 \author{Sarai Dery Folkestad}
 \affiliation[NTNU]{Department of Chemistry, Norwegian University of Science and Technology, NO-7491 Trondheim, Norway}
 \author{Sonia Coriani}
 \email{soco@kemi.dtu.dk}
 \affiliation[DTU]{DTU Chemistry, Technical University of Denmark, DK-2800 Kongens Lyngby, Denmark}
 \alsoaffiliation[NTNU]{Department of Chemistry, Norwegian University of Science and Technology, NO-7491 Trondheim, Norway}

\begin{document}

\begin{abstract}
We present an implementation of a damped response framework for calculating resonant inelastic X-ray scattering (RIXS) at the equation-of-motion coupled cluster singles and doubles (CCSD) 
and 
second-order approximate coupled cluster singles and doubles (CC2)
levels of theory in the 
open-source program $e^T$. This framework lays the foundation for future extension to higher excitation methods (notably, the coupled cluster singles and doubles with perturbative triples, CC3) and to multilevel approaches. 

Our implementation adopts a fully relaxed ground state,
and different variants of the core-valence separation projection technique to address convergence issues.
Illustrative results are compared with those obtained within the frozen-core core-valence separated approach, available in 
Q-Chem, as well as with experiment.

The performance of the CC2 method is evaluated in comparison with that of CCSD. It is found that, while the CC2 method is noticeably inferior to CCSD for X-ray absorption spectra, the quality of the CC2 RIXS spectra is often comparable to that of the CCSD level of theory, when the same valence excited states are probed. Finally, we present preliminary RIXS results for a solvated molecule in aqueous solution.
\end{abstract}

\section{\label{sec:Intro}Introduction}
X-ray spectroscopy has gained tremendous popularity over the past decades.\cite{mobilio_14, lamberti_16, bergmann_17, nisoli_17} This is a result of massive investments in the development of improved light sources and the refinement of detection techniques, both at large scale
experimental facilities and at table-top laboratory setups.\cite{mobilio_14, bergmann_17} 
Experiments such as X-ray absorption (XAS)\cite{lamberti_16, bergmann_17, nisoli_17} and X-ray emission (XES)\cite{lamberti_16} are nowadays routinely carried out, in particular in the soft X-ray regime (0.1-5keV), where, e.g., the $K$-edge (1$s^{-1}$) of lighter elements like carbon, nitrogen and oxygen, the
$L$-edge ($2p^{-1}$) of third row elements like sulfur, chlorine and silicon, and the $M$-edge ($3d^{-1}$)  of transition metals are probed.
Moreover, thanks to the X-ray pulse sources nowadays available from free-electron lasers and high-harmonic generation setups, it has become possible to use X-ray techniques to monitor ultrafast molecular transformations with 
 unprecedented temporal, spatial and energetic resolution.~\cite{Leone:XRayUltraFastRev:18,Leone:UltrafastChemRx:2018}

\begin{figure}
    \centering
    \includegraphics[scale=0.5]{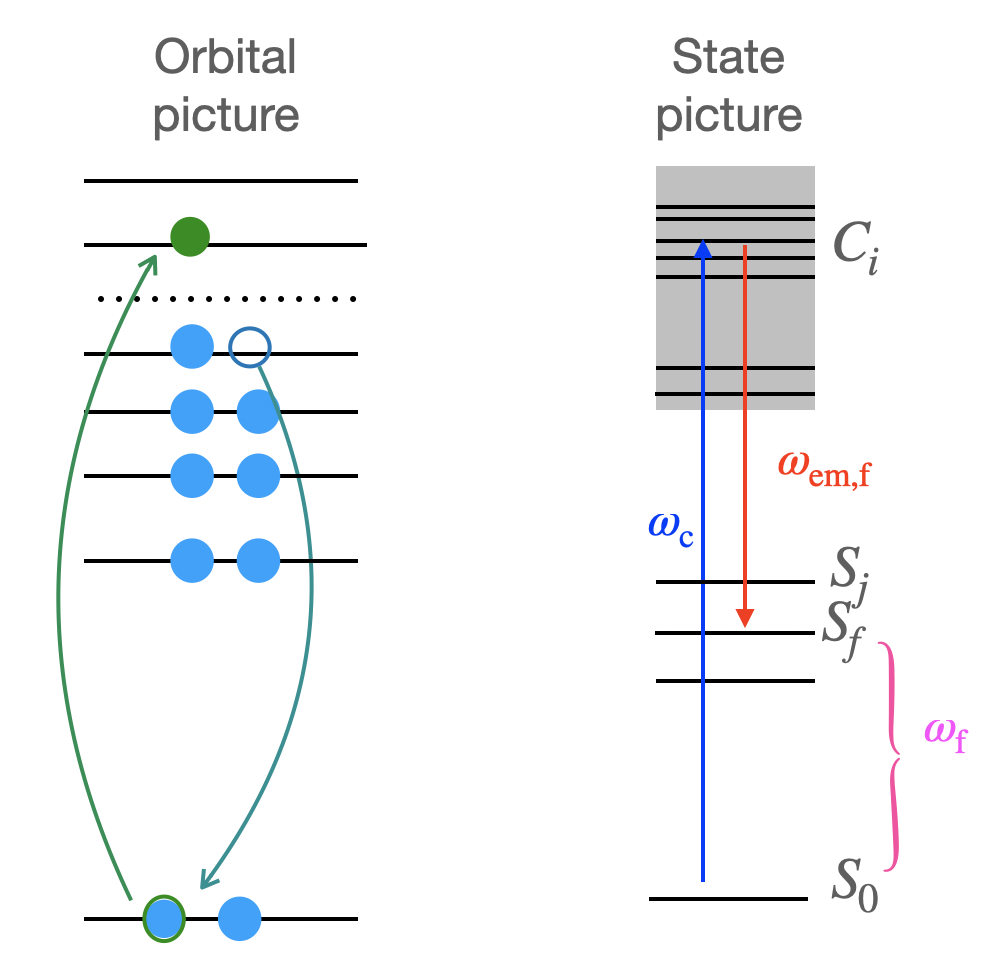}%{Figures/RIXS_Scheme.png}
    \caption{Orbital and state schematics of the RIXS process. {\em Left}: A core electron is excited to a virtual orbital, and a valence electron falls down to fill the core hole. 
    {\em Right}: The system is excited into core state $C_i$ (XAS), then decays into valence excited state $S_f$ (XES), emitting photons of energy $\omega_{\textrm{em,f}}$, lower than that used 
    in the XAS step.}
    \label{fig:RIXS_Scheme}
\end{figure}

An X-ray technique, that has significantly benefited from the recent advances in X-ray light sources and instrumentation,
in particular synchrotrons, is resonant inelastic X-ray scattering (RIXS). This technique is also known as resonant 
(radiative) 
inelastic 
X-ray Raman scattering.\cite{gelmukhanov_99,kotani_01,ament_11,eckert_17,TRIXS}
Broadly speaking, RIXS can be regarded as a two-photon process, equivalent to
vibrational Raman scattering.
The first step in both processes is the absorption of a photon, for RIXS in the X-ray region (i.e., XAS),
and for conventional Raman in the near-IR to UV region.\cite{Miedema17}
Subsequently, another photon is emitted in the same energy region, i.e., XES for RIXS.\cite{gelmukhanov_99,carra_95}
Thus, where conventional Raman is used to probe vibrational energy states,
RIXS probes (valence) electronic excited states.
Like its 
(infrared-visible) analogue, this two-photon process in the X-ray region is complementary to simple absorption spectroscopy, since their selection rules differ. 
In terms of molecular orbitals, one can also say that, while XAS probes unoccupied valence orbitals, RIXS studies the correlation between occupied and unoccupied valence orbitals. 
Figure~\ref{fig:RIXS_Scheme}
schematically illustrates the RIXS process both in terms of orbital excitations/de-excitations and of transitions between electronic states. An important difference of RIXS compared to valence region Raman spectroscopy is its element specificity due to the localized nature of the core orbitals, which is seen for all X-ray spectroscopies.\cite{Schmitt_14}

With RIXS experiments becoming more feasible and popular during the last decade, 
even in time-resolved regimes,~\cite{TRIXS} the interest in theoretical approaches capable of simulating the RIXS observables has also increased, as it is deemed essential for the interpretation of the experimental spectra.~\cite{RIXS:Besley,CPP:RIXS:ADC,2019_jctc_Faber_RIXS,2020_pccp_Faber_CVS-CPP,2020_pccp_Nanda_RIXS,2020_Nanda_JCP,RIXS:TDDFT:1,RIXS:TDDFT:2,PhysRevB.99.104306,PhysRevB.103.115136}
A non-exhaustive list in this regard includes methods based on the 
algebraic diagrammatic construction (ADC) approach,\cite{CPP:RIXS:ADC} on time-dependent density functional theory (TD-DFT),\cite{RIXS:Besley,RIXS:TDDFT:1,RIXS:TDDFT:2} on
multi-configurational self-consistent field
(MCSCF)~\cite{RIXS:Odelius} and density-matrix renormalization group (DMRG) theories,\cite{RIXS:DMRG1}
and on coupled cluster (CC) theory.\cite{2019_jctc_Faber_RIXS,2020_pccp_Nanda_RIXS,2020_pccp_Faber_CVS-CPP,2020_Nanda_JCP}

The key quantity when computing the spectra is the RIXS cross section,~\cite{gelmukhanov_99} based on the Kramers-Heisenberg-Dirac (KHD) formula.~\cite{KramersHeisenberg,Dirac1,Dirac2} In this formulation the amplitudes are given in terms of a sum-over-states (SOS) expression. It has been argued\cite{Long:Raman,ament_11,CPP:RIXS:ADC} that the SOS formula may be slowly convergent and 
require a large number of terms in order to span the spectral range of valence and core-excited states. A closed-form expression would then be preferable.
One such expression was derived and implemented
within the ADC framework by Rehn et al. in 2017.\cite{CPP:RIXS:ADC}
Two years later, Faber and Coriani\cite{2019_jctc_Faber_RIXS} proposed an analogous theoretical framework for computing RIXS spectra 
(along with XES) within coupled cluster singles and doubles (CCSD) response theory. Illustrative spectra for small molecular systems were there presented based on a prototyping python implementation.\cite{2019_jctc_Faber_RIXS}
A similar approach was later developed for frozen-core equation-of-motion (fc-EOM-)CCSD by Nanda et al.,\cite{2020_pccp_Nanda_RIXS} and implemented in the commercial code Q-Chem.\cite{Qchem541}
Even though CCSD is not considered the gold standard within the framework of single reference wavefunctions, it provides a computationally feasible approach that often yields sufficiently accurate results.\cite{Izak_20} The method is therefore of interest, since more accurate approaches, like, e.g., coupled cluster singles and doubles with a perturbative treatment of triple excitations, CCSD(T) and CC3,\cite{CC3,CC3resp,CC3_new} often advocated as the gold standards,
become computationally too demanding with the generally available resources, when considering molecules of a certain size.\cite{Izak_20} In the future, though, it would be of interest to extend the framework also to these methods.
%While coupled cluster singles and doubles methods with a perturbative treatment of triple excitations, i.e. CCSD(T) and CC3,\cite{CC3,CC3resp,CC3_new}are often considered the golden standard for calculations using single reference wavefunctions,\cite{Izak_20} these methods are still  computationally too demanding, when considering molecules of a certain size, with the generally available computational resources. CCSD however, provides a computationally cheaper approach that still provides high accuracy results.\cite{Izak_20}

When computing molecular properties within a coupled cluster wavefunction ansatz, one can choose either the response (RSP) framework~\cite{CC:RSP:1990,ijqc-ove,WaveFuncRev} or the more approximate equation-of-motion (EOM) framework.~\cite{Stanton:93:EOMCC,bartlett_07,Krylov:EOMRev:07,coriani-cc-ci} The latter is often the preferred choice, as it yields results of similar accuracy as RSP at a (slightly) lower cost.\cite{caricato_09} 
The two formulations are equivalent for the exact wave function (full CC limit). 
 For truncated CC parametrizations, the excitation energies determined with the two methods are the same. Oscillator strengths (and other transition strengths in general) differ,
 as the EOM strengths are not size-intensive,
 \cite{koch_94,caricato_09,WaveFuncRev}  i.e.,
 the error of the EOM methods scales with the size of the system. Fortunately, however, the error remains small until hundreds of correlated electron pairs are considered.\cite{caricato_09}As for the (non-linear) RIXS scattering cross sections considered in this work, the still limited number of previous studies has shown only small differences between EOM and RSP based results.\cite{2019_jctc_Faber_RIXS}

Despite the results presented by Faber and Coriani,\cite{2019_jctc_Faber_RIXS,2020_pccp_Faber_CVS-CPP} and by 
Nanda et al.\cite{2020_pccp_Nanda_RIXS,2020_Nanda_JCP} appearing promising, available implementations of RIXS at the CC level are to date  still scarce. To the best of our knowledge, none is available in an open-source code. In this study, we present such an implementation in the $e^T$ code,\cite{eT} which will be available in a future release. The $e^T$ code features a very efficient implementation of the Cholesky decomposition\cite{beebe_77,ROEGGEN_86,koch_03,folkestad_19} of the two-electron integrals, which allows for cost-effective implementations in quantum chemistry. Unlike the approach of Refs.~\citenum{2020_pccp_Nanda_RIXS} and \citenum{2020_Nanda_JCP}, we here work within the framework of a fully relaxed ground state and only invoke the frozen-core approximation  when computing the valence excited states. In other words, the ground state is obtained with all electrons correlated, whereas the valence excited states are computed by excluding core excitations from the excitation manifold. The core excited states are conversely calculated by projecting out pure valence excitations from the excitation manifold.\cite{coriani_15,coriani_15_erratum}
Furthermore, the core-valence-separation\cite{Cederbaum1980} (CVS) projection within the damped response solver can be turned-off if desired, although it is generally recommended to use it for convergence.\cite{2020_pccp_Faber_CVS-CPP} In addition, two flavours of the CVS projector are available, namely ``full''  CVS\cite{coriani_15,coriani_15_erratum} and CVS with uncoupled valence singles (CVS-uS).\cite{2020_Nanda_JCP}

We will begin by briefly reviewing the relevant theoretical background for the new implementation in Section \ref{sec:theory}, followed by a summary of the computational details in Section \ref{sec:details}.
Numerical results are presented and discussed in Section \ref{Results}. Specifically, to
 validate our implementation, the RIXS and XES spectra of H$_2$O were calculated at the EOM-CCSD and EOM-CC2 levels of theory and compared to results obtained using the fc-CVS-EOM-CCSD approach, as well as experiment (see Section \ref{sec:water}). Furthermore, the performance of both levels of theory for RIXS calculations was evaluated by simulating the RIXS spectra of methanol (CH$_3$OH) and comparing with experiment. The RIXS spectra of H$_2$S and para-nitro-aniline (PNA) were likewise evaluated, even though experimental data for comparisons are not available (see Section \ref{sec:cc2secs}). Finally, in Section \ref{sec:imidazole}, the CCSD implementation was employed to simulate imidazole in an aqueous solution.
Concluding remarks are given at the end.

\section{Quantities of interest and implementation details}\label{sec:theory}
%All equations on which the current implementation is based can be found in Refs. \citenum{CPP:Kauczor:CC} and \citenum{2019_jctc_Faber_RIXS}. They will therefore not as a rule be repeated here.

The main quantity of RIXS is the RIXS cross section defined as\cite{CPP:RIXS:ADC,2019_jctc_Faber_RIXS,gelmukhanov_94,gelmukhanov_99,ament_11}
\begin{align}
\begin{split}
    \sigma^{0f}_{\theta}&=  \frac{\omega_{\textrm{em,f}}}{\omega_{\textrm{c}}}\frac{1}{15}\sum_{XY}\Bigg\{
    \Big(2-\frac{1}{2}\sin^2\theta\Big)
    \mathcal{F}_{XY}^{0f}(\omega_{\textrm{c}})\mathcal{F}_{XY}^{f0}(\omega_{\textrm{c}})\\
    &+\Big(\frac{3}{4}\sin^2\theta-\frac{1}{2}\Big)
    \left[\mathcal{F}_{XX}^{0f}(\omega_{\textrm{c}})\mathcal{F}_{YY}^{f0}(\omega_c)+\mathcal{F}_{YY}^{0f}(\omega_{\textrm{c}})\mathcal{F}_{XX}^{f0}(\omega_{\textrm{c}})\right]
    \Bigg\}~,
    \end{split}\label{eq:rixs_cross}
\end{align}
where $\theta$ is the angle between the incident and emitted photon and $X$ and $Y$ indicate two components of the electric dipole moment operator, i.e.,   $\hat{X}$ and $\hat{Y}$. Furthermore, as also shown in Fig. \ref{fig:RIXS_Scheme},  $\omega_{\textrm{c}}$ is the (core) resonance energy and $\omega_{\textrm{em,f}}$ is the emission energy associated with the final valence excited state, $f$, i.e., \mbox{$\omega_{\textrm{em,f}}=(\omega_{\textrm{c}}-\omega_{\textrm{f}})$}. The quantities $\mathcal{F}_{XY}^{0f}(\omega_{\textrm{c}})$ and $\mathcal{F}_{XY}^{f0}(\omega_{\textrm{c}})$ are the RIXS transition moments between the initial state, $0$, and the final valence excited state, $f$. These transition moments can be calculated based on the SOS expression for the KHD amplitudes,\cite{gelmukhanov_99,Long:Raman,CPP:RIXS:ADC} which reads
\begin{align}
\label{generalF}
    \mathcal{F}^{f0}_{XY}(\omega_{\textrm{c}})=\sum_{n}\left[
    \frac{\langle\Psi_f|\hat{X}|\Psi_n\rangle \langle\Psi_n|\hat{Y}|\Psi_0\rangle}{\omega_n-(\omega_{\textrm{c}}+i\gamma_n)}
    +\frac{\langle\Psi_f|\hat{Y}|\Psi_n\rangle \langle\Psi_n|\hat{X}|\Psi_0\rangle}{\omega_n+(\omega_{\textrm{em,f}}+i\gamma_n)}
    \right],
\end{align}
where $|\Psi_n\rangle$ is the wavefunction for the excited state $n$,  $\gamma_n$ is the corresponding inverse lifetime, and $i$ is the imaginary phase. In order to simplify Eq.~\eqref{generalF}, one can assume that all excited states have the same inverse lifetime $\gamma$, resulting in
\begin{align}
\label{KHD-gamma}
    \mathcal{F}^{f0}_{XY}(\omega_{\textrm{c}})=\sum_{n}\left[
    \frac{\langle\Psi_f|\hat{X}|\Psi_n\rangle \langle\Psi_n|\hat{Y}|\Psi_0\rangle}{\omega_n-(\omega_{\textrm{c}}+i\gamma)}
    +\frac{\langle\Psi_f|\hat{Y}|\Psi_n\rangle \langle\Psi_n|\hat{X}|\Psi_0\rangle}{\omega_n+(\omega_{\textrm{em,f}}+i\gamma)}\right].
\end{align}
As discussed elsewhere,~\cite{CPP:RIXS:ADC,2019_jctc_Faber_RIXS}
this expression corresponds to a two-photon transition matrix element where a damping factor is added to the frequencies $\omega_\textrm{c}$ and $\omega_{\textrm{em,f}}$ in order to maintain the resonant
condition \mbox{$\omega_{\textrm{c}}-\omega_f = \omega_{\textrm{em,f}}$}.

Our implementation of the KHD amplitude
in Eq.~\eqref{KHD-gamma}
is based on a CC ground-state wavefunction, i.e., on the exponential ansatz\cite{helgaker2014molecular} $|\Psi^{\textrm{CC}}_0\rangle = \exp(T)|0\rangle$. Here, $|0\rangle$ is the reference (Hartree-Fock) wavefunction and $T=\sum_m t_m\hat{\tau}_m$ is the cluster operator, consisting of the excitation operators $\hat{\tau}_m$, acting as $\hat{\tau}_m|0\rangle = |m\rangle$, and of the coupled cluster amplitudes, $t_m$. 

For the CCSD wavefunction considered here, only single and double excitations are included in the cluster operator. The EOM-CC formalism~\cite{Stanton:93:EOMCC,Krylov:EOMRev:07} is moreover employed to formally parametrize the final state wavefunction $\Psi_f$ and the intermediate state wavefunctions $\Psi_n$.

Following the derivation of \citeauthor{2019_jctc_Faber_RIXS},\cite{2019_jctc_Faber_RIXS}
the EOM-CC right and left RIXS transition moments can be computed as
\begin{align}
\begin{split}
    \mathcal{F}_{XY}^{0f}(\omega_{\textrm{c}})&=\Big[
     \bar{t}^X(\omega_{\textrm{em,f}}-i\gamma)\textbf{A}^Y+\bar{t}^Y(-\omega_{\textrm{c}}+i\gamma)
     \textbf{A}^X\\
    &-(\bar{t}\cdot\xi^X)\bar{t}^Y(-\omega_{\textrm{c}}+i\gamma)
    -(\bar{t}\cdot\xi^Y)\bar{t}^X(\omega_{\textrm{em,f}}-i\gamma)\Big] R_f\\
    &+(\bar{t}\cdot R_f)\Big[\bar{t}^Y(-\omega_{\textrm{c}}+i\gamma)\cdot \xi^X
    +\bar{t}^X(\omega_{\textrm{em,f}}-i\gamma)
    \cdot
    \xi^Y
    \Big]
    \end{split}\label{eq:rixs_transmom1}\\
    \begin{split}
    \mathcal{F}_{XY}^{f0}(\omega_{\textrm{c}})&=
    -L_f\Big\{
    \textbf{A}^X t^Y(\omega_{\textrm{c}}+i\gamma)+\textbf{A}^Y
    t^X(-\omega_{\textrm{em}}-i\gamma)\\
    &-(\bar{t}\cdot\xi^X)
    t^Y(\omega_{\textrm{c}}+i\gamma)
    -(\bar{t}\cdot\xi^Y)
    t^X(-\omega_{\textrm{em,f}}-i\gamma)\\
    &-(\bar{t}\cdot t^Y(\omega_{\textrm{c}}+i\gamma))\xi^X
    -(\bar{t} \cdot t^X(-\omega_{\textrm{em,f}}-i\gamma))\xi^Y
    \Big\}.
    \end{split}\label{eq:rixs_transmom2}
\end{align}
In the equations above, $\bar{t}$ is the vector containing the CC ground-state multipliers, and $L_f$  and $R_f$ are the left and right EOM-CC eigenvectors of state $f$, respectively. In addition, $t^X(\omega+i\gamma)$ and $\bar{t}^X(\omega-i\gamma)$ are the vectors containing the CC response amplitudes and multipliers, respectively, while $\textbf{A}^X$ is the EOM property Jacobian
defined, for instance, in Eq.~(20) of Ref.~\citenum{2019_jctc_Faber_RIXS}. The RHS vectors 
$\xi^X$ and $\eta^X$ can be found, for instance,
in Eqs.~(9) and~(18) of Ref.~\citenum{2019_jctc_Faber_RIXS}, respectively.

Our implementation of the damped response solver, used to determine the response amplitudes and multipliers, is based on the algorithm presented in Refs.~\citenum{2019_jctc_Faber_RIXS} and \citenum{2013_jcp_Kauczor_CPP}. We will therefore not repeat it here, but rather refer the reader to those publications for details.
In essence, the complex response amplitudes and multipliers are decomposed into real and  imaginary components, e.g.,  $t^X(\omega + i\gamma) = t^X_{\Re}(\omega + i\gamma) + i t_{\Im}^X(\omega + i\gamma)$. The response equations are then recast in a (real) pseudosymmetric form, like
\begin{align}
    \label{Matrix_form}
    \left(\begin{matrix} % or pmatrix or bmatrix or Bmatrix or ...
        ({\bf{A}} - {\omega}  {\bf{I}}) &   \gamma {\bf{I}} \\
        \gamma {\bf{I}}& -({\bf{A}} - {\omega}  {\bf{I}}) \\
    \end{matrix}\right)
    \left(\begin{matrix} % or pmatrix or bmatrix or Bmatrix or ...
         {{t}}^X_\Re \\
         {{t}}^X_\Im\
    \end{matrix}\right) =
    \left(\begin{matrix} % or pmatrix or bmatrix or Bmatrix or ...
        - \xi^X_\Re \\
        \xi^X_\Im\\
    \end{matrix}\right)
\end{align}
and solved using an iterative subspace algorithm.~\cite{2019_jctc_Faber_RIXS,2013_jcp_Kauczor_CPP}

Note that within the EOM-CC framework, the solution of the left damped response equation, 
yielding the response multipliers, is strictly decoupled from the solution of the right damped equation (Eq.~\eqref{Matrix_form}). This is because the EOM right-hand-side (RHS) vector $\eta^X$ does not contain contributions from the response amplitudes. This is not the case in CC linear response (LR-CC) theory, where the response amplitudes enter the RHS vectors of the response multiplier equations.\cite{2019_jctc_Faber_RIXS} This implies that both RHS vectors of the EOM response equations, $\xi^X$ or $\eta^X$, 
are either strictly real or strictly imaginary, only depending on the nature of the operator $\hat{X}$.
Since the damped response amplitudes and multipliers are decomposed into real and  imaginary components, the real and imaginary parts of the transition moments in Eqs.~\eqref{eq:rixs_transmom1}-\eqref{eq:rixs_transmom2} can be evaluated separately.

% \begin{align}
% \begin{split}
%     \mathcal{F}_{\Re, XY}^{0f}(\omega)&=\Big[
%      \bar{t}^X(\omega'-i\gamma)A^Y+\bar{t}^Y(-\omega+i\gamma)A^X\\
%     &-(\bar{t}\cdot\xi^X)\bar{t}^Y(-\omega+i\gamma)
%     -(\bar{t}\cdot\xi^Y)\bar{t}^X(\omega'-i\gamma)\Big]R_f\\
%     &+(\bar{t}R_f)\Big[\bar{t}^Y(-\omega+i\gamma)\xi^X
%     +\bar{t}^X(\omega'-i\gamma)\xi^Y
%     \Big]
%     \end{split}\\
%     \begin{split}
%     \mathcal{F}_{\Im, XY}^{0f}(\omega)&=\Big[
%      \bar{t}^X(\omega'-i\gamma)A^Y+\bar{t}^Y(-\omega+i\gamma)A^X\\
%     &-(\bar{t}\cdot\xi^X)\bar{t}^Y(-\omega+i\gamma)
%     -(\bar{t}\cdot\xi^Y)\bar{t}^X(\omega'-i\gamma)\Big]R_f\\
%     &+(\bar{t}R_f)\Big[\bar{t}^Y(-\omega+i\gamma)\xi^X
%     +\bar{t}^X(\omega'-i\gamma)\xi^Y
%     \Big]
%     \end{split}\\
%     \begin{split}
%     \mathcal{F}_{\Re, XY}^{f0}(\omega)&=
%     -L_f\Big[
%     A^Xt_{\Re}^Y(\omega+i\gamma)+A^Yt_{\Re}^X(-\omega'-i\gamma)\\
%     &-(\bar{t}\cdot\xi^X_{\Re})t^Y_{\Re}(\omega+i\gamma)
%     -(\bar{t}\cdot\xi^Y_{\Re})t^X_{\Re}(-\omega'-i\gamma)
%     \\
%     &-(\bar{t}t^Y_{\Re}(\omega+i\gamma))\xi^X_{\Re}
%     -(\bar{t}t^X_{\Re}(-\omega'-i\gamma))\xi^Y_{\Re}
%     \Big]
%     \end{split}
%     \\
%     \begin{split}
%     \mathcal{F}_{\Im, XY}^{f0}(\omega)&=
%     -L_f\Big[
%     A^Xt^Y_{\Im}(\omega+i\gamma)+A^Yt^X_{\Im}(-\omega'-i\gamma)\\
%     &-(\bar{t}\cdot\xi^X_{\Re})t^Y_{\Im}(\omega+i\gamma)
%     -(\bar{t}\cdot\xi^Y_{\Re})t^X_{\Im}(-\omega'-i\gamma)\\
%     &-(\bar{t}t^Y_{\Im}(\omega+i\gamma))\xi^X_{\Re}
%     -(\bar{t}t^X_{\Im}(-\omega'-i\gamma))\xi^Y_{\Re}
%     \Big]
%     \end{split}
% \end{align}
In RIXS, we consider only components of the real electric dipole moment operators, $\hat{X}$, and thus $\xi^X_{\Im}$ and 
$\eta^X_{\Im}$ are zero. Furthermore, it is the real part of the RIXS transition strengths that is of interest when computing intensities of the spectra. Thus, we only need to evaluate the real part of the transition moment products, e.g., $\mathcal{F}^{0f}_{XY}(\omega_c)\mathcal{F}^{f0}_{ZU}(\omega_c)$. 
Moreover, since the focus of our implementation has been on CCSD, the current damped response solver for CC2 uses the full singles and doubles vectors. This will be optimized in the future. Algorithm \ref{alg:rixs} summarizes the main steps in the calculation of the RIXS cross section.
 \begin{algorithm}
 \begin{algorithmic}[1]
\State{Calculate (or input) the resonant core excitation energy, $\omega_{\textrm{c}}$, using a Davidson solver and CVS.}
\State{Calculate valence excitation energies, $\omega_{f}$, and solution vectors, $L_f$ and $R_f$, using a Davidson solver in a space orthogonal to the pure core space.}
\State{Determine emission energies $\omega_{\textrm{em,f}}=\omega_{\textrm{c}}-\omega_{f}$.}
\State{Collect $\omega_{\textrm{c}}$ and $\omega_{\textrm{em}}$ in array $\omega$ and generate $-\omega$.}
\For {sign($\omega$)}
\For {$X = x$-,$y$-,$z$-component of the dipole operator}
 \State{Solve right damped response equations for the response amplitudes, $t^X(\omega+i\gamma)$ and $t^X(-\omega-i\gamma)$.}
 \State{Solve left damped response equations for the response multipliers, $\bar{t}^X(-\omega+i\gamma)$ and $\bar{t}^X(\omega-i\gamma)$.}
\EndFor
\EndFor
\For{f = 1, $N_{\textrm{valence\_excitations}}$}
\State Calculate transition moments, $\mathcal{F}_{XY}^{0f}(\omega_{\textrm{c}})$ and $\mathcal{F}_{XY}^{f0}(\omega_{\textrm{c}})$ given in Eqs. \eqref{eq:rixs_transmom1} and \eqref{eq:rixs_transmom2}. 
% \begin{align}
% \begin{split}
%     \mathcal{F}_{XY}^{0f}(\omega_{\textrm{c}})&=\Big[
%      \bar{t}^X(\omega_{\textrm{em,f}}-i\gamma)A^Y+\bar{t}^Y(-\omega_{\textrm{c}}+i\gamma)A^X\\
%     &-(\bar{t}\cdot\xi^X)\bar{t}^Y(-\omega_{\textrm{c}}+i\gamma)
%     -(\bar{t}\cdot\xi^Y)\bar{t}^X(\omega_{\textrm{em,f}}-i\gamma)\Big]R_f\\
%     &+(\bar{t}R_f)\Big[\bar{t}^Y(-\omega_{\textrm{c}}+i\gamma)\xi^X
%     +\bar{t}^X(\omega_{\textrm{em,f}}-i\gamma)\xi^Y
%     \Big]
%     \end{split}\\
%     \begin{split}
%     \mathcal{F}_{XY}^{f0}(\omega_{\textrm{c}})&=
%     -L_f\Big[
%     A^Xt^Y(\omega_{\textrm{c}}+i\gamma)+A^Yt^X(-\omega_{\textrm{em}}-i\gamma)\\
%     &-(\bar{t}\cdot\xi^X)t^Y(\omega_{\textrm{c}}+i\gamma)
%     -(\bar{t}\cdot\xi^Y)t^X(-\omega_{\textrm{em,f}}-i\gamma)\\
%     &-(\bar{t}t^Y(\omega_{\textrm{c}}+i\gamma))\xi^X
%     -(\bar{t}t^X(-\omega_{\textrm{em,f}}-i\gamma))\xi^Y
%     \Big]
%     \end{split}
% \end{align}
\For{$\theta=\textrm{angle}_1,\textrm{angle}_M$}
\State {Calculate RIXS cross section, $\sigma^{0f}_{\theta}$, for the angle, $\theta$, between the incident and scattered photon given in Eq. \eqref{eq:rixs_cross}.}
% \begin{align}
% \begin{split}
%     \sigma^{0f}_{\theta}&=  \frac{\omega_{\textrm{em,f}}}{\omega_{\textrm{c}}}\frac{1}{15}\sum_{XY}\Bigg[
%     \Big(2-\frac{1}{2}\sin^2\theta\Big)
%     \mathcal{F}_{XY}^{0f}(\omega_{\textrm{c}})\mathcal{F}_{XY}^{f0}(\omega_{\textrm{c}})\\
%     &+\Big(\frac{3}{4}\sin^2\theta-\frac{1}{2}\Big)
%     (\mathcal{F}_{XX}^{0f}(\omega_{\textrm{c}})\mathcal{F}_{YY}^{f0}(\omega)+\mathcal{F}_{YY}^{0f}(\omega_{\textrm{c}})\mathcal{F}_{XX}^{f0}(\omega_{\textrm{c}}))
%     \Bigg]
% \end{split}
% \end{align}
\EndFor
\EndFor
 \end{algorithmic}
 \caption{
 Schematic illustration of the RIXS implementation.}
 \label{alg:rixs}
 \end{algorithm}
% \\

 Our current implementation of RIXS enables the user to either input a value for the core excitation energy, or to request a certain core excitation to be calculated as the initial step (see Algorithm \ref{alg:rixs}). 
 As an additional feature, one can switch on a projection in the damped response solver. 
 It has previously been observed that larger systems, or even small systems described with larger basis sets, display a poor convergence in the damped response solver.\cite{2020_pccp_Faber_CVS-CPP, 2020_Nanda_JCP} This, however, can be remedied by applying a CVS projector in the solver.\cite{2020_pccp_Faber_CVS-CPP, 2020_Nanda_JCP} 
 \citeauthor{2020_pccp_Faber_CVS-CPP}\cite{2020_pccp_Faber_CVS-CPP} noted that when simply considering the convergence, this projection needs only be employed when solving the response amplitude equations with positive frequencies and the response multiplier equations with negative frequencies. Our implementation of the damped response solver includes the projection in this manner. Additionally, it was reported by \citeauthor{2020_Nanda_JCP}
 \cite{2020_Nanda_JCP} that when considering systems with significant charge transfer character, the core and valence transitions cannot be completely decoupled. Indeed, the valence single excitations were shown to couple significantly to the core space, which then led the authors to formulate a different variant of CVS, named CVS-uS.\cite{2020_Nanda_JCP} In CVS-uS, only the double valence excitations were excluded from the solution of the damped response equations. This procedure was found not to impede the convergence of the damped response equations, while it was shown to improve the description of the spectra. CVS-uS has also been implemented as a projector in $e^T$ and can be invoked if desired. Note, however, that \citeauthor{2020_Nanda_JCP}\cite{2020_Nanda_JCP} employed the CVS-uS scheme when solving for all damped response equations, and not only for the potentially divergent ones. This is also the case for their implementation of the ``full'' CVS (denoted CVS-0) in Q-Chem. In the following, we have chosen to apply the CVS and CVS-uS projectors primarily to improve convergence of the potentially divergent response equations. An in-depth investigation of the differences of the results following the two schemes is postponed to future studies. A schematic illustration of our implementation of the two projections can be found in Algorithm \ref{alg:cvs}.\\
 \begin{algorithm}[htbp!]
 \begin{algorithmic}[1]
\If { (CVS requested \textbf{and} (right equations \textbf{and} positive $\omega$) \textbf{or} (left equations \textbf{and} negative $\omega$))}
\For{i=1,$N_{\textrm{excitations}}$}
\If{i does not involve the core orbital of interest}
\State vector$_i$ = 0.0
\EndIf
\EndFor
\Else {\textbf{if} (CVS-uS requested \textbf{and} (right equations \textbf{and} positive $\omega$) \textbf{or} (left equations \textbf{and} negative $\omega$))}
\For{i=1,$N_{\textrm{double\_excitations}}$}
\If{i does not involve the core orbital of interest}
\State vector$_i$ = 0.0
\EndIf
\EndFor
\EndIf
 \end{algorithmic}
 \caption{Schematic illustration of our implementation of the CVS and CVS-uS projectors on a vector. The projection is only invoked for the right (amplitude) response equations when the input frequency is positive, and for the left (multiplier) response equations when the frequency is negative.}\label{alg:cvs}
 \end{algorithm}

Finally, the calculation of nonresonant X-ray emission spectra (XES) was also implemented, following the approach outlined by \citeauthor{2019_jctc_Faber_RIXS}.\cite{2019_jctc_Faber_RIXS} Accordingly, we compute the emission energy as the difference between the ionization energy of the core ionized state, c, and the
valence ionized state, v
\begin{equation}
    E_{\textrm{em}}= \textrm{IE}^{\textrm{c}} - 
    \textrm{IE}^{\textrm{v}}.
\end{equation}
Ionization energies and ionization vectors are obtained in $e^T$ as excitations into a bath orbital.\cite{stanton_99,coriani_15,moitra_22}
The intensities are based on the (EOM) dipole transition moments
\begin{align}
    T_{X}^{\textrm{vc}} = (L_{\textrm{v}}\textbf{A}^{X}R_{\textrm{c}})-(\bar{t}\cdot R_\textrm{c})(L_\textrm{v} \cdot \eta^{X})-(L_\textrm{v}\textbf{I}R_\textrm{c})(\bar{t}\cdot \eta^{X}).\label{eq:xes}
\end{align}
A schematic illustration of the XES implementation can be found in Algorithm \ref{alg:xes}.\\
 \begin{algorithm}
 \begin{algorithmic}[1]
\State{ Calculate the desired core ionization energy, IE$^{\textrm{c}}$, and solution vectors, $L_{\textrm{c}}$ and $R_{\textrm{c}}$, using a Davidson solver and CVS}
\State{ Calculate valence ionization energies, IE$^{\textrm{v}}$, and solution vectors, $L_{\textrm{v}}$ and $R_{\textrm{v}}$, using a Davidson solver in a space orthogonal to the pure core space}
\For{v = 1, $N_{\textrm{valence\_excitations}}$}
\For {$X$ = $x$-,$y$-,$z$-component of the dipole operator}
\State{Determine transition moments, $T^{\textrm{vc}}_X$ and $T^{\textrm{cv}}_X$} given in Eq. \eqref{eq:xes}.
\State {Determine diagonal element of the transition strength, $S_{XX}^{\textrm{vc}}=T_{X}^{\textrm{vc}}T_{X}^{\textrm{cv}}$}
\State {Add $X$ component contribution to XES oscillator strength, $f^{\textrm{osc}}_{\textrm{vc}}$}
\begin{align*}
    f^{\textrm{osc}}_\textrm{vc}~+\!=
    \frac{2}{3}(\textrm{IE}^{\textrm{c}}-
    \textrm{IE}^{\textrm{v}})S_{XX}^\textrm{vc}
\end{align*}
\EndFor
\EndFor
 \end{algorithmic}
 \caption{Schematic illustration of the XES implementation. \\
 Note that XES calculations do not require the use of the damped response solver.}\label{alg:xes}
 \end{algorithm}
\section{Computational Details}
\label{sec:details}
All calculations have been performed with a development version\cite{development_version_eT} of the open source program $e^T$ (Ref.~ \citenum{eT}). All core and valence excitation spectra have been determined with a regular Davidson solver. The core excitations were computed in a space orthogonal to the pure valence one, and vice versa for the valence excitations.~\cite{coriani_15} The RIXS spectra have been determined by using our damped response solver with the CVS-uS and CVS projection. For H$_2$O and H$_2$S the damped response calculations have also been performed without any CVS projection. In addition, RIXS calculations for H$_2$O have been carried out at the fc-CVS-0-EOM-CCSD level of theory using Q-Chem.~\cite{Qchem541}
For H$_2$O, CH$_3$OH, H$_2$S and PNA calculations were performed both at the EOM-CCSD and EOM-CC2 levels of theory. For imidazole in aqueous solution, the calculations were only performed at the EOM-CCSD level of theory. The 6-311++G** basis\cite{krishnan1980a,clark1983a} with additional Rydberg functions (3$s$3$p$) was employed in the case of H$_2$O (available in the supplementary information, textformat file), whereas for CH$_3$OH we adopted the \mbox{aug-cc-pVTZ} basis.\cite{dunning1989a,kendall1992a} For the remaining systems, the 6-311++G** basis set was used. 
Convergence thresholds for the response calculations were:  $10^{-6}$ for CH$_3$OH, PNA and imidazole in H$_2$O, $10^{-8}$ for H$_2$O to compare with the study by \citeauthor{2020_pccp_Nanda_RIXS},\cite{2020_pccp_Nanda_RIXS} and $2\cdot10^{-8}$ for H$_2$S as the RIXS cross sections here are very small. All RIXS calculations employed a damping factor of 0.0045563 a.u. and the RIXS cross sections are shown for a scattering angle of 45$^{\circ}$.

% As the $e^T$ program does not employ symmetry, all excitation energies were also calculated with the Dalton program\cite{daltonpaper,Dalton2020} in order to obtain the symmetries of the transitions.

The geometries of H$_2$O and H$_2$S were determined from experimental parameters,\cite{HOY19791,COOK1975237} while the structure of PNA was taken from \citeauthor{2020_Nanda_JCP}.\cite{2020_Nanda_JCP} The geometries of CH$_3$OH and of imidazole in H$_2$O were optimized at the MP2/cc-pCVTZ level of theory. 
The cartesian coordinates of all systems can be found in the supplementary information (textformat file).

The Mulliken symmetry notation\cite{Mulliken_55} has been employed throughout unless otherwise stated.

All calculations were carried out on the DTU HPC resources.\cite{DTU_DCC_resource}

%%%%%%%%%%%%%%%%%%%%%
\section{Results and discussion}\label{Results}
\subsection{Validation}\label{sec:water}
%%%%SONIA 11.14
\subsubsection{H$_2$O}
%\revS{SONIA}
%\\
H$_2$O was used to validate our approach by comparing different available CC implementations for the calculation of RIXS spectra. 

\begin{figure}
     \centering
     \includegraphics[width=0.9\textwidth]{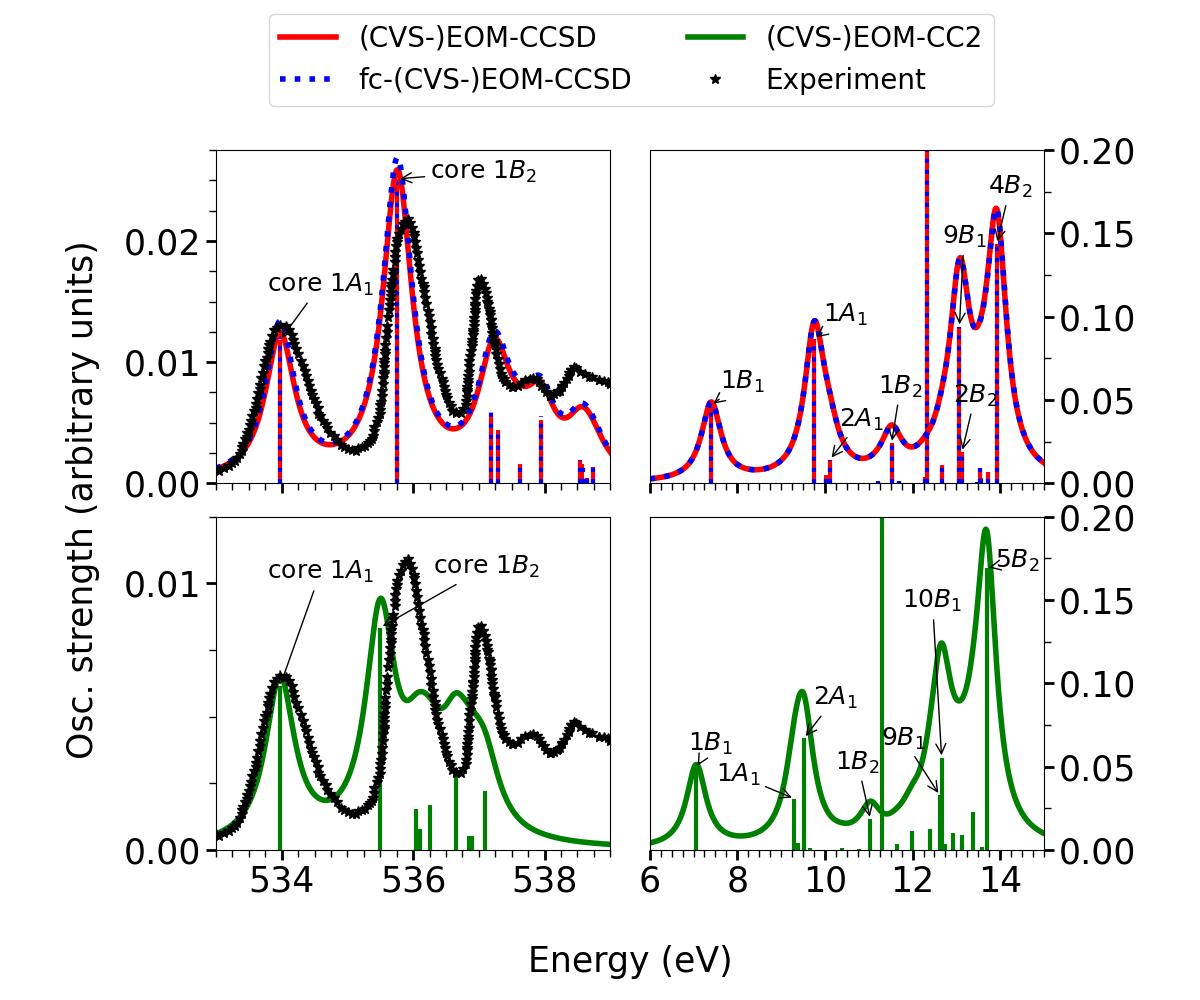}%{Figures/water_xas_and_val.png}
     \caption{H$_2$O: XAS spectra at the O $K$-edge (left) and valence absorption spectra in a space orthogonal to the O core space (right) calculated using different methods. For all methods, 10 core transitions were determined and plotted. For CC2, 33 valence transitions were computed and plotted, while only 30 transitions were considered for CCSD.
     The 6-311++G** basis with additional Rydberg functions was used for all calculations. Experimental data (black) was digitized from \citeauthor{schirmer_93}\cite{schirmer_93} The spectra were simulated by applying a Lorentzian broadening with a half width at half maximum (HWHM) of 0.27~eV. The calculated XAS spectra were shifted to align with the first experimental band by 
     $-$0.52~eV for CVS-EOM-CC2, $-$1.72~eV for CVS-EOM-CCSD and 
     $-$1.24~eV for fc-CVS-EOM-CCSD. The first valence ionization energy is shown as a vertical line spanning the entire intensity range. Arrows
indicate the main transitions probed. Note that $1A_1$ labels the first valence excitation of $A_1$ symmetry.}
     \label{fig:water_xas}
 \end{figure}

As the RIXS calculations rely on the preliminary computation of the core transition, corresponding to the core resonance of interest, as well as of the valence transitions, the XAS and \mbox{UV-Visible} absorption spectra are also shown. Moreover, EOM-CC2 and EOM-CCSD results from $e^T$ and fc-EOM-CCSD results from Q-Chem are compared. The XAS and valence spectra can be seen in 
Fig.~\ref{fig:water_xas}. The underlying raw data (energies and oscillator strengths) used to build the spectra can be found in Tables \ref{SI-tab:xas_water} and \ref{SI-tab:val_water} in the SI.
Table~\ref{tab:ips} collects the calculated ionization energy (IE) thresholds of all systems considered in this study.
\setlength{\tabcolsep}{10pt} % Default value: 6pt
\begin{table}[h]
    \caption{First valence and core ionization energy thresholds calculated 
    with $e^T$. 
    Core ionization energies were obtained using the CVS projector to span a space orthogonal to the valence excitation space. Valence ionization energies were computed in a space orthogonal to the core space of interest. The core space considered is given in parenthesis next to the computed values.}
    \label{tab:ips}
    \centering
    \begin{tabular}{llcc}
    \hline
Molecule&Method & IE/eV & Core IE/eV \\\hline
\multirow{4}{*}{H$_2$O}
&EOM-CC2  & 11.30(O) & 538.09(O)\\
&EOM-CCSD & 12.33(O) & 541.46(O)\\
&fc-EOM-CCSD & 12.31(O) & 540.98(O) \\
&Experiment\cite{SNOW199049,SANKARI2003647}&12.65&539.79(O)\\\hline
\multirow{3}{*}{CH$_3$OH}
&EOM-CC2 & 10.05(C);10.06(O)&293.07(C); 536.98(O)\\
&EOM-CCSD & 11.02(C);11.03(O)&
293.18(C); 540.43(O)\\
&Experiment\cite{tao_92,drake_77,PELLEGRIN2020112258}&10.85&292.42(C); 539.16(O)\\\hline
\multirow{3}{*}{H$_2$S}&EOM-CC2 & 9.77(S) & 2473.89(S)\\
&EOM-CCSD & 10.00(S) & 2475.72(S)\\
&Experiment\cite{PREST1983315,CARROLL1987281}&10.46&2478.32(S)\\\hline
\multirow{3}{*}{PNA}
&EOM-CC2 & 7.8641(C) 
&
291.58(C)\\
%291.5799\\
&EOM-CCSD & 8.34(C) &292.31(C)\\
&Experiment\cite{johnstone_73,turci_96}&8.43&291.1(C)
\\\hline
\multirow{2}{*}{Imidazole in H$_2$O}
&EOM-CCSD & 8.78(N) &406.41(N)\\
&Experiment\cite{tentscher_15,nolting_08}&8.51&403.9(N)\\\hline
    \end{tabular}
\end{table}

As seen from Fig.~\ref{fig:water_xas} (and Table~\ref{SI-tab:xas_water}), the main differences between the CVS-EOM-CCSD and the fc-CVS-EOM-CCSD XAS results are a smaller shift in energy of the latter to align with experiment, and minimally larger oscillator strengths, as already reported in previous studies.\cite{vidal_19_xas} 
The differences between the CCSD and CC2 spectra, on the other hand, are much larger. Apart from the first two transitions, the CC2 core transitions are noticeably more closely spaced and their oscillator strengths are much smaller. 
As a result, where the CCSD spectra show reasonably good agreement with the experimental XAS spectrum in terms of both relative peak positions and intensities, the EOM-CC2 spectrum above 535.5 eV is compressed because of the too small separation between the peaks. 
This behaviour has already been reported in previous studies.~\cite{Lanczos1,CARBONE_19}
It is a known fact that CC2 does not describe XAS to a fully satisfactory degree (see, e.g., Refs. \citenum{frati_19} and \citenum{reinholdt_21}).

Considering the valence excitation spectrum, see right panel of Fig.~\ref{fig:water_xas} and Table 
\ref{SI-tab:val_water} in the SI,
we observe that the only difference between the EOM-CCSD and fc-EOM-CCSD spectra is a small (0.02 eV) red shift of the latter with respect to the former. Aside from a slightly larger red shift, the CC2 UV spectrum corresponds well with the CCSD spectra with only small variations above the calculated ionization threshold. The CC2 ionization threshold is, however, obtained at a much lower energy (1 eV) than the CCSD one.

Despite the similar spectral shapes, closer inspection of the
energies and intensities (tabulated in Table~\ref{SI-tab:val_water}) reveals differences between CC2 and CCSD in the energetic ordering and symmetry 
of some of the excited states contributing to the peaks above 10 eV. This may affect the appearance of RIXS spectral slices in the low emission energy region.
Also note that in our notation $1A_1$ is the first valence excited state of $A_1$ symmetry, since the ground state is not part of the analysis. 

\begin{figure}
    \centering
    \includegraphics[width=\textwidth]{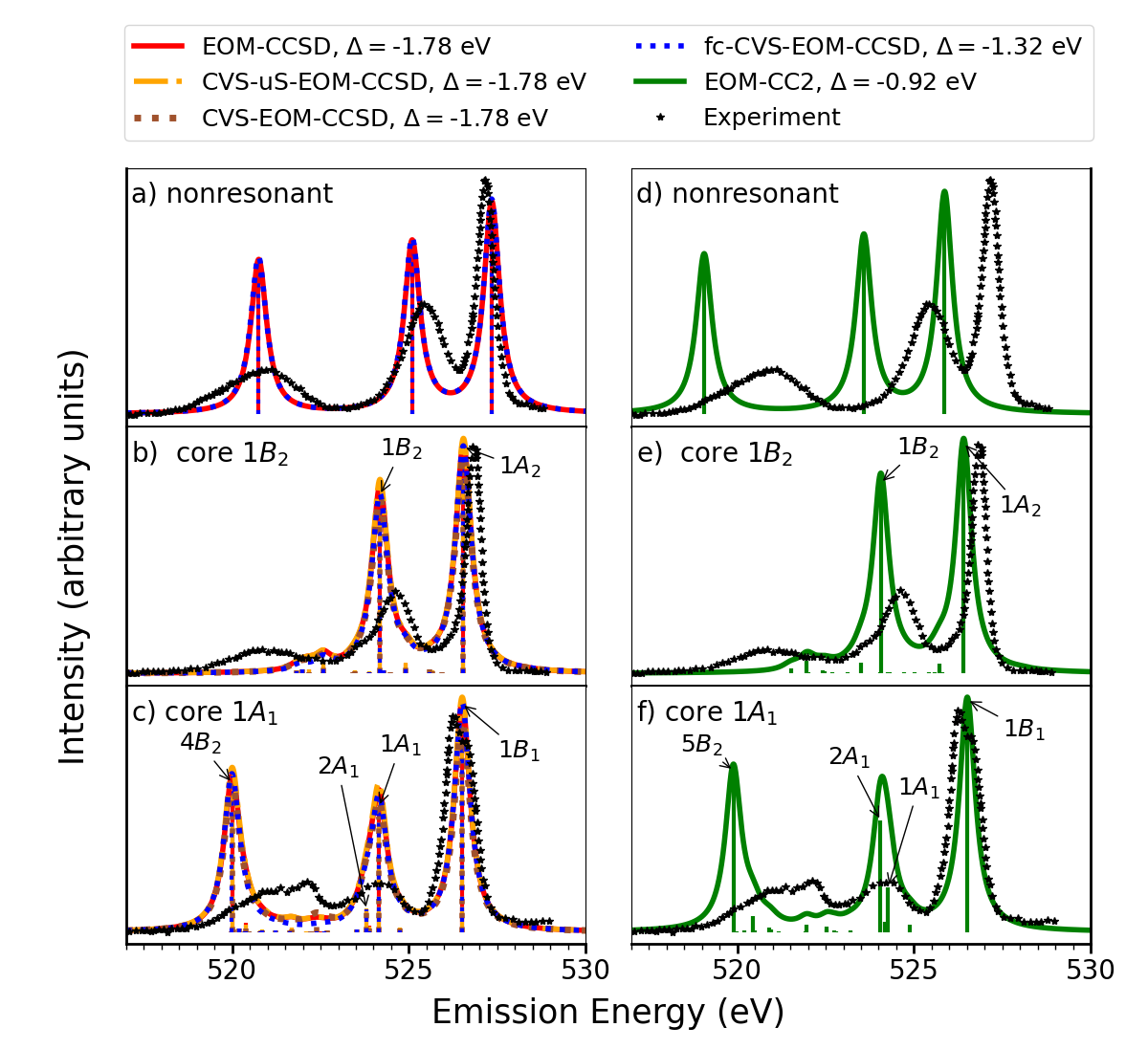}%{Figures/new_water_RIXS.png}
    \caption{H$_2$O: Nonresonant XES (a and d) and RIXS spectra at resonance with the energy of the first (c and f) and second (b and e) core excitation at the O $K$-edge. The 6-311++G** basis with additional Rydberg functions was used for all calculations. The spectra were simulated by Lorentzian broadening with HWHM=0.27~eV. Experimental data was digitized from \citeauthor{weinhardt_2012}\cite{weinhardt_2012} No projector was employed in the damped response solver to obtain the EOM-CCSD and EOM-CC2 results, while the CVS and CVS-uS results were obtained by applying the CVS projectors to remove all valence excitations 
    and all pure valence double excitations, respectively. 
     Arrows indicate the main transitions probed. Note that $1A_1$ labels the first valence excitation of $A_1$ symmetry.
    Computed spectra were shifted by the amounts indicated in the legend, which correspond to shifts required to align to the high energy peak in the experimental RIXS spectra at the core $1A_1$ resonance.
    }
    \label{fig:water_rixs_all}
\end{figure}

We now study the RIXS spectra (spectral slices) for the first two core resonances, $1s\to4a_1$ (total symmetry $A_1$) and $1s\to2b_2$ (total symmetry $B_2$), as well as the nonresonant XES spectra.
% In this work we choose to adhere to Mulliken notation unless otherwise explicitly stated.\\
%
The RIXS spectra are given in panels b, c, e and f of
Fig.~\ref{fig:water_rixs_all}, and the nonresonant XES spectra in panels a and d.
As before, freezing the core orbitals in the ground state merely results in a shift compared to calculations based on a fully relaxed ground state. 
Also, for this molecule, using either the CVS projector or the CVS-uS projector when solving for the damped response vectors, does not affect the resulting spectra (see Figs.~\ref{fig:water_rixs_all}b and \ref{fig:water_rixs_all}c). This was also reported by \citeauthor{2020_Nanda_JCP},\cite{2020_Nanda_JCP} who noted that differences between the two schemes do not always arise, and in particular not for very small systems and/or small basis sets. 

When comparing the CC2 and CCSD  RIXS results, one immediately notes the very similar spectral profiles at both core resonances, even though the overall
intensity of the CC2 spectrum is systematically lower.
The profiles are also in reasonably good agreement with the experiment.
%
%At both levels of theory, the intensities (relative to the first band)
%of the second and third band at the first core resonance (panels c and f) and of the second band at the  second core resonance (panels b and e) are overestimated. Moreover, the third band of  RIXS spectrum at the core 1$A_1$ resonance is peaking at too low emission energy, as also previously reported.\cite{2020_pccp_Faber_CVS-CPP}
%AK suggestion for above paragraph:
At both levels of theory, the intensities 
of the two low energy bands at the first core resonance are overestimated relative to the intensity of the high energy band (panels c and f). This is also observed at the second core resonance (panels b and e). Moreover, the third band of the RIXS spectrum at the core 1$A_1$ resonance is calculated to peak at a too low emission energy, as also previously reported.\cite{2020_pccp_Faber_CVS-CPP}

Analysing which specific transitions contribute to the RIXS spectra 
(see also Table \ref{tab:water_rixs_assign_res} and Section \ref{SI-sec:h2o}, Tables \ref{SI-tab:val_water}-\ref{SI-tab:water_rixs_2nd} in the SI)
at the core 1$A_1$ resonance,
we find that the band at around 526.5 eV is only due to the emission to the $1B_1$ valence state.
The next band is dominated by the emission into the $1A_1$ and $2A_1$ valence states.
These two states have opposite energetic ordering and reversed RIXS intensities at the CC2 and CCSD level, but closer inspection of their natural transition orbitals 
(see Fig. \ref{SI-fig:water_NTOs}) 
reveals that the character of the $1A_1$ CCSD state is the same as the character of the $2A_1$ CC2 state. Likewise, the character of the $2A_1$ CCSD state
is the same at that of the $1A_1$ CC2 state.
The characters of the $1B_1$, $1A_2$ and $1B_2$ states are the same for the two methods.
Thus, states of same character at both levels of theory have comparable (relative) RIXS intensities. A similar situation occurs for the third band at around 520 eV. Here, the dominant contribution is due to emission to the $4B_2$ state at the CCSD level and to 
the  $5B_2$ at the CC2 level, but the two states have the same electronic character.

In general, as we move towards lower emission energy, more and more of the electronic states probed by RIXS  are highly energetic (inner valence) states, which fall above the first ionization threshold. Whether the CCSD and CC2 methods are still capable of yielding a reliable description of these states becomes arguable, in particular if the states have a significant contribution of double excitations. An inaccurate description of the inner valence electronic states could be one of the reasons for the drift in energy and overestimated relative intensity of the low emission energy band in the RIXS spectrum of the core $1A_1$ resonance.  Additional studies at a higher level of theory are nonetheless needed to put this hypothesis on firmer grounds. 
%Both methods predict} %CCSD predicts 
%similar intensities of the \revAK{two low energy} band\revAK{s}, somewhat similar to what is seen in the experiment. %CC2 on the other hand yields a much lower intensity of the third feature. In fact, the specific excited state probed at the CCSD level of theory, dominating the third RIXS peak (4$B_2$), is not amongst those reached at the CC2 level. Instead, it occurs at much higher valence excitation energy. 

As for the RIXS spectrum at resonance with the second core excitation (core $1B_2$), we again observe a good correspondence between methods.
The two strongest spectral features originate from the emission into the $1A_2$ and $1B_2$ valence excited states, which have the same characters for the CCSD and CC2 methods.
Also, the theoretically predicted spectra are in reasonable agreement with experiment. 
We do observe a small additional shift 
(with  respect to experiment), though, which can be explained by the shift in energy of the theoretically predicted second core excitation compared to experiment. As the energy difference between the first and second core excitation is theoretically underestimated, in particular for CC2, 
the computed RIXS emission energies will likewise be underestimated. Thus, applying the shift calculated based on the RIXS spectra at the first core resonance for all core resonances will result in the theoretically predicted spectra at the second core resonance to fall at sligthly  too low emission energies.

To summarize, we find that the calculations at the CC2 level, while not nearly as adequate for simulating XAS as CCSD, yield results of a comparable quality for RIXS, when the same valence excited states (i.e., states of the same electronic character) are probed by the two methods.

The most noticeable differences between CCSD  and CC2 is observed in the nonresonant XES spectrum, see panels a and d of Figure~\ref{fig:water_rixs_all}. More precisely, while the spectral profiles are still very similar, the CC2 XES spectrum is visibly shifted in energy.
This is clearly a consequence of the lower core ionization energies found for CC2.
\begin{table}[h]
\caption{H$_2$O: Overview of the main valence states probed in RIXS by the different methods at the first and second core resonance (core 1$A_1$ and core 1$B_2$, respectively). As the results with EOM-CCSD, EOM-CVS-CCSD and EOM-CVS-uS-CCSD are practically identical, the table only shows one set of results. The 6-311++G** basis with additional Rydberg functions was used for all calculations. Reported energies are not shifted, as they are in Fig. \ref{fig:water_rixs_all}.  Note that $1A_1$ here denotes the first valence excitation of $A_1$ symmetry and not the ground state.}
    \label{tab:water_rixs_assign_res}
    \centering
    \begin{tabular}{cccccc}
    \hline
    \multicolumn{2}{c}{EOM-CCSD}&\multicolumn{2}{c}{fc-CVS-0-EOM-CCSD}&\multicolumn{2}{c}{EOM-CC2}\\
         Energy/ eV& Probed state & Energy/ eV& Probed state  & Energy/ eV& Probed state\\\hline
         \multicolumn{6}{c}{core 1$A_1$ resonance}\\\hline
         528.2906& 1$B_1$& 527.8331& 1$B_1$ &527.4307&1$B_1$\\
         525.9389 & 1$A_1$&525.4821 & 1$A_1$&525.1898 &1$A_1$\\
         525.5838 & 2$A_1$ &525.1229 &2$A_1$& 524.9754&2$A_1$\\
         521.7740 & 4$B_2$&521.3106&4$B_2$&520.8069&5$B_2$%521.3625&12$B_1$
         \\\hline
          \multicolumn{6}{c}{core 1$B_2$ resonance}\\\hline
          528.3254& 1$A_2$& 527.8658 &1$A_2$&527.3286 &1$A_2$\\
          525.9615 & 1$B_2$&525.5011&1$B_2$ &524.9822&1$B_2$
          \\\hline
    \end{tabular}
\end{table}
\FloatBarrier
\subsection{Evaluation of performance of CC2 and CCSD for RIXS and XES}\label{sec:cc2secs}
\subsubsection{CH$_3$OH}
The nonresonant emission and RIXS of methanol have been the subject of several studies in the past.~\cite{Larkins_1990,benkert_16,vaz_da_cruz_2019} In particular, a combined experimental and theoretical (based on density functional theory) study was carried out in 2016,\cite{benkert_16} and a thorough computational investigation at the second-order ADC level of theory was published in 2019.\cite{vaz_da_cruz_2019} 
We use the molecule here to further evaluate the relative performance of CC2 and CCSD.

%As for water, while the computed valence spectra look similar, the XAS spectra show large 
%discrepancies between the EOM-CCSD and EOM-CC2 methods (see Section \ref{SI-sec:ch3oh}). 
%\revS{!OBS!! ARE WE SURE? The XAS do look rather similar too.}

The CC2 and CCSD valence and XAS spectra, reported Section \ref{SI-sec:ch3oh},
are similar, even though the CC2 XAS spectrum at the oxygen $K$-edge is slightly compressed 
and less intense compared to the CCSD ones. This was also found for water. 
The first two peaks in the oxygen $K$-edge XAS spectrum corresponds to core excited states of $A'$ symmetry at both levels of theory, and they are in the following labelled ``core 1$A'$'' and
``core 2$A'$'' to distinguish them from the valence ones. 
Likewise, at the carbon $K$-edge, the first peak corresponds to a core excited state of $A'$ symmetry.
As before, we use ``$1A^\prime$'' to indicate the first valence excited state of $A^\prime$ symmetry, and not the ground state.

\begin{figure}
    \centering
    \includegraphics[width=\textwidth]{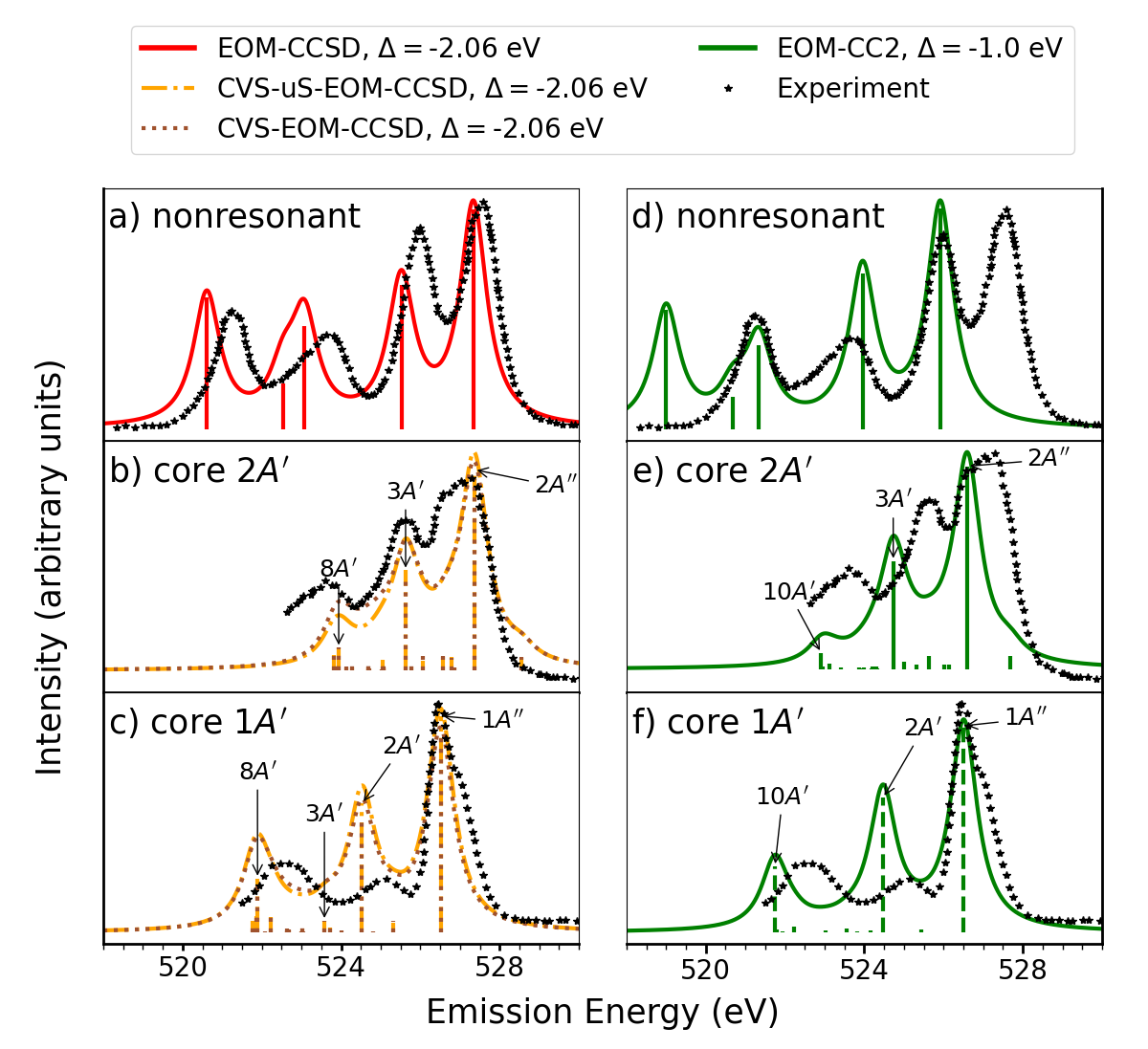}%{Figures/new_methanol_rixs_O.png}
    \caption{CH$_3$OH: Nonresonant XES (a and d) and RIXS spectra at resonance with the first (c and f) and second (b and e) core excitation at the O $K$-edge. All spectra were calculated using the aug-cc-pVTZ basis set. They were broadened by a Lorentzian broadening with HWHM=0.41~eV. Experimental data was digitized from \citeauthor{benkert_16}\cite{benkert_16} No projections have been employed in the damped response solver to obtain EOM-CC2 results, while the CVS and CVS-uS results were obtained by applying a CVS projection removing all valence excitations and all pure valence double excitations, respectively. 
    Arrows indicate the main valence transitions probed. Note that $1A^{\prime}$ denotes the first valence excitation of $A^{\prime}$ symmetry. Computed spectra were shifted by the amounts indicated in the legend to align to the high energy peak in the experimental RIXS spectra at the first core resonance.
    }
    \label{fig:methanol_rixs_O}
\end{figure}

\begin{figure}
    \centering
    \includegraphics[width=\textwidth]{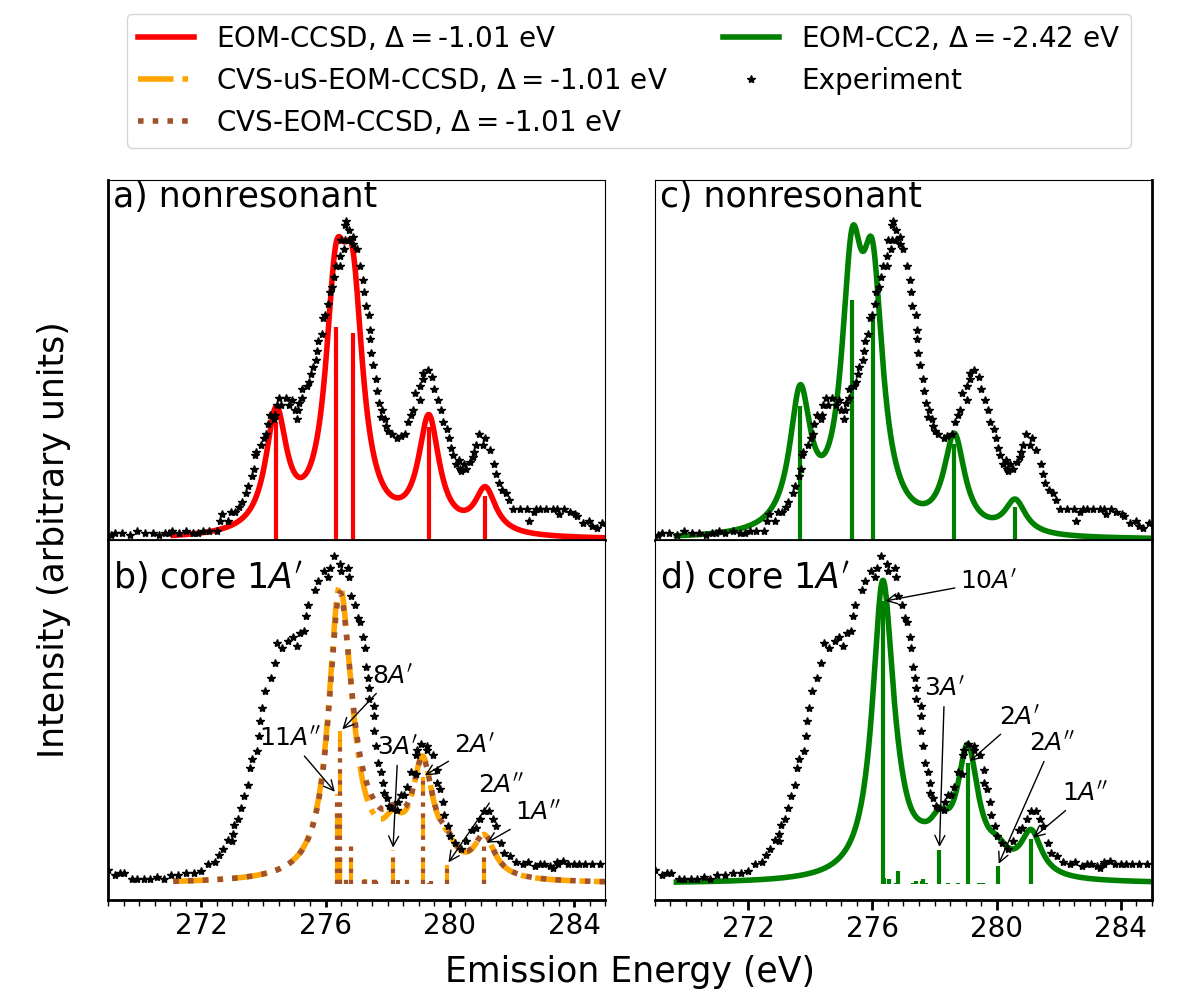}%{Figures/new_methanol_rixs_C.png}
    \caption{CH$_3$OH: Nonresonant (a and c) and RIXS spectra at resonance with the energy of the first core excitation (b and d) at the C $K$-edge. All spectra are computed with the aug-cc-pVTZ basis set. A Lorentzian broadening has been applied with HWHM=0.41eV. Experimental data was digitized from \citeauthor{benkert_16}\cite{benkert_16} No projections have been employed in the damped response solver to obtain EOM-CCSD and EOM-CC2 results, while the CVS and CVS-uS projections have been used in the this solver to obtain the remaining results. Arrows indicate the main valence transitions probed. Note that $1A^{\prime}$ denotes the first valence excitation of $A^{\prime}$ symmetry. The spectra were shifted by the amounts indicated in the legend to align to the high energy peak of the experimental RIXS spectra at the first core resonance.
    }
    \label{fig:methanol_rixs_C}
\end{figure}

The RIXS spectra have been computed at resonance with the first and second core excitation at the oxygen $K$-edge (unshifted values are 534.0 eV and 535.2 eV for EOM-CC2 and 535.5 eV and 537.5 eV for EOM-CCSD), and with the first core excitation at the carbon $K$-edge (unshifted values are 290.1 eV for EOM-CC2 and 289.0 eV for EOM-CCSD). 
Nonresonant X-ray emission spectra at both edges have also been computed. 
The spectra are shown in Figs. \ref{fig:methanol_rixs_O}a-f and \ref{fig:methanol_rixs_C}a-d. Note that only the emission energy regions above 521 eV for O and 276 eV for C were reached in the RIXS spectra calculations with the number of valence excited states here considered.

At the oxygen $K$-edge (Fig. \ref{fig:methanol_rixs_O}) the CVS-uS-EOM-CCSD and CVS-EOM-CCSD spectra are basically identical, and the EOM-CC2 results are in good qualitative agreement with the CCSD ones with respect to the spectral shape. 
It is, however, noted that the intensities predicted by EOM-CC2 are less than 50\% of those predicted with the CCSD-based methods. 
As also seen previously, 
after shifting all spectra by the value needed to align to the highest energy peak in the core $1A^\prime$ RIXS spectrum, the EOM-CC2 spectra are  more shifted compared to experiment than the CCSD ones. This is particularly evident at the second core resonance, where the compressed XAS
%core excitation 
spectrum at the CC2 level of theory results in a significant shift of the CC2 RIXS spectrum compared to experiment (see Figs. \ref{fig:methanol_rixs_O}b and \ref{fig:methanol_rixs_O}e). 

The symmetry and electronic character of the electronic states corresponding to the two most intense peaks are the same at the CC2 and CCSD levels of theory, see Tables \ref{tab:met_rixs_assign_resO} and \ref{tab:met_rixs_assign_resC}, 
and Fig. \ref{SI-fig:methanol_NTOs} for the natural transition orbitals.
%The transitions corresponding to the two most intense peaks are the same \revAK{with respect to symmetry label }at the CC2 and CCSD levels (see Tables \ref{tab:met_rixs_assign_resO} and \ref{tab:met_rixs_assign_resC}). 

%The characters of the probed transitions at the CCSD level of theroy is found to correspond to those probed at the CC2 level of theory with the same symmetry labels.
Furthermore, the CCSD $8A'$ state, which is the main contributor to the third peak in the CCSD RIXS spectrum, 
has the same electronic character as the CC2 $10A'$ state (see Fig. \ref{SI-fig:methanol_NTOs}),
which yields the intensity to the third peak in the CC2 RIXS spectrum. 
Thus, as in the case of water,
%, while CCSD probes an additional transition compared to CC2, 
%SONIA: which one?
the electronic states probed by RIXS at the CC2 level of theory are the same as obtained at the CCSD level of theory.

\begin{table}[h]
\caption{CH$_3$OH: Overview of the main valence states probed in RIXS with the different methods at the first and second core resonance at the O $K$-edge (core 1$A'$ and core 2$A'$). All calculations used the aug-cc-pVTZ basis set. Note that $1A^{\prime}$ denotes the first valence excitation of $A^{\prime}$ symmetry.}
    \label{tab:met_rixs_assign_resO}
    \centering
    \begin{tabular}{cccccc}
    \hline
 \multicolumn{2}{c}{CVS-uS-EOM-CCSD}&\multicolumn{2}{c}{CVS-EOM-CCSD}&\multicolumn{2}{c}{EOM-CC2}\\
                     Energy/ eV& Probed state & Energy/ eV& Probed state  & Energy/ eV& Probed state\\
    \hline
    \multicolumn{6}{c}{core $1A'$ resonance}\\\hline
         528.5730& 1$A''$& 528.5730& 1$A''$ &527.5061&1$A''$\\
         526.5896 & 2$A'$&526.5896 & 2$A'$&525.4704 &2$A'$\\
         525.6423 & 3$A'$ &525.6423 &3$A'$& -&-\\
         523.9544 & 8$A'$&523.9544&8$A'$&522.7419&10$A'$
         \\\hline
         \multicolumn{6}{c}{core 2$A'$ resonance}\\\hline
         529.4225& 2$A''$& 529.4225& 2$A''$ &527.5958 &2$A''$\\
         527.6905  & 3$A'$&527.6905  & 3$A'$&525.7272 &3$A'$\\
         526.0026 & 8$A'$ &526.0026 &8$A'$& 523.9087&10$A'$
         \\\hline
    \end{tabular}
\end{table}
\begin{table}[h]
\caption{CH$_3$OH: Overview of the main valence states probed in RIXS with the different methods at the first core resonance at the C $K$-edge (core 1$A'$). All calculations used the aug-cc-pVTZ basis set. Note that $1A^{\prime}$ denotes the first valence excitation of $A^{\prime}$ symmetry.}
    \label{tab:met_rixs_assign_resC}
    \centering
    \begin{tabular}{cccccc}
    \hline
    \multicolumn{2}{c}{CVS-uS-EOM-CCSD}&\multicolumn{2}{c}{CVS-EOM-CCSD}&\multicolumn{2}{c}{EOM-CC2}\\
         Energy/ eV& Probed state & Energy/ eV& Probed state  & Energy/ eV& Probed state\\\hline
         \multicolumn{6}{c}{core 1$A'$ resonance}\\\hline
         282.1232& 1$A''$& 282.1232& 1$A''$ &283.5314&1$A''$\\
        280.9191&2$A''$&280.9191&2$A''$&282.4503&2$A''$\\
        280.1359&2$A'$&280.1359&2$A'$&281.4935&2$A'$\\
        279.1827&3$A'$&279.1827&3$A'$&280.5784&3$A'$\\
        277.4937&8$A'$&277.4937&8$A'$&278.7536&10$A'$\\
        277.3721&11$A''$&277.3721&11$A''$&-&-
         \\\hline
    \end{tabular}
\end{table}

Also at the carbon $K$-edge, see Fig.~\ref{fig:methanol_rixs_C}, 
the CC2 and CCSD spectra show very similar spectral profiles. 
The most intense transitions probed at resonance with the core 1$A'$-state are found to be $1A''$,
$2A'$ and $10A'$ for EOM-CC2, while according to the EOM-CCSD methods the $1A''$, $2A'$, $8A'$ and
$11A''$ states are the main contributors to the RIXS spectrum. As discussed above the CCSD $8A'$ state corresponds to the CC2 $10A'$ state.
It is thus observed that, with the given number of valence excited states considered, the most intense peak at the EOM-CC2 level of theory is ascribed to a single transition, whereas it is described by two different transitions at the CVS-EOM-CCSD and CVS-uS-EOM-CCSD levels of theory. However, it is evident that more valence excited states should be considered at both levels of theory to ensure that this peak is described to a satisfactory degree.

Finally, as observed for water, the CC2 and CCSD XES spectra at both $K$-edges have almost identical profiles, 
but EOM-CC2 spectra are much more shifted from experiment than the EOM-CCSD ones, due to the gross underestimation of the core ionization energy by CC2.

\FloatBarrier
\subsubsection{H$_2$S}
As another showcase molecule, we consider H$_2$S. Experimental XAS spectra for this molecule have been reported by \citeauthor{BODEUR1985}\cite{BODEUR1985}
and by \citeauthor{Reynaud_1996},\cite{Reynaud_1996} while an experimental $1s3p$ RIXS map was reported as a conference communication by 
\citeauthor{Kavcic_2009}\cite{Kavcic_2009} 
A computational benchmark study of XAS of sulfur containing molecules of interest in astrochemistry, here including H$_2$S, has been conducted by \citeauthor{Bilalbegovic2021436}\cite{Bilalbegovic2021436} Furthermore, \citeauthor{Ertan_H2S}\cite{Ertan_H2S} computed both potential energy surfaces for core excited states and RIXS spectra for the first two core resonances. These spectra however focus on the energy region where the S2$p^{-1}$ excited states are probed, which is beyond the scope of our study. Additionally, the core excitations considered are very close in energy and might thus be excited simultaneously. In this study, we investigate the RIXS spectra for the core excitations of largest intensity for the three peaks visible in the experimental XAS spectrum. For both low energy peaks two close-lying transitions occur and therefore both are investigated.

Also in this case, the valence spectra are in good agreement between CC2 and CCSD methods, while the XAS spectra differ somewhat, albeit here the difference is small (see Section \ref{SI-sec:h2s}). The core resonances to consider are the close-lying core~$1B_2$ and core~$1A_1$, as well as the close-lying core~$1A_2$ and core~$1B_1$. Finally, the resonance corresponding to the third peak in the XAS spectrum is core $3B_2$. This is the case at both levels of theory. 

As mentioned, RIXS spectra have been computed at the resonance of the most intense transitions for the three peaks of the XAS (unshifted values are 2468.6 eV, 2468.8 eV, 2471.4 eV, 2471.6 eV and 2473.0 eV at the EOM-CC2 level of theory and 2469.8 eV, 2470.3 eV, 2473.3 eV, 2473.5 eV and 2475.2 eV at the EOM-CCSD level of theory). These are all below the calculated core ionization threshold at the CCSD level of theory, while the third experimental feature is above the threshold at the CC2 level (see Table~\ref{tab:ips}). 
Nonresonant emission spectra were computed as well. As for H$_2$O, three different strategies were adopted to compute the CCSD RIXS spectra: no projection was applied during the solution of the (damped) response equations (labeled EOM-CCSD), a CVS projector was applied to remove all valence excitations (labeled CVS-EOM), and 
a CVS projector was applied to remove all pure valence double excitations (labeled CVS-uS-EOM).

\begin{figure}
    \centering
    \includegraphics[width=\textwidth]{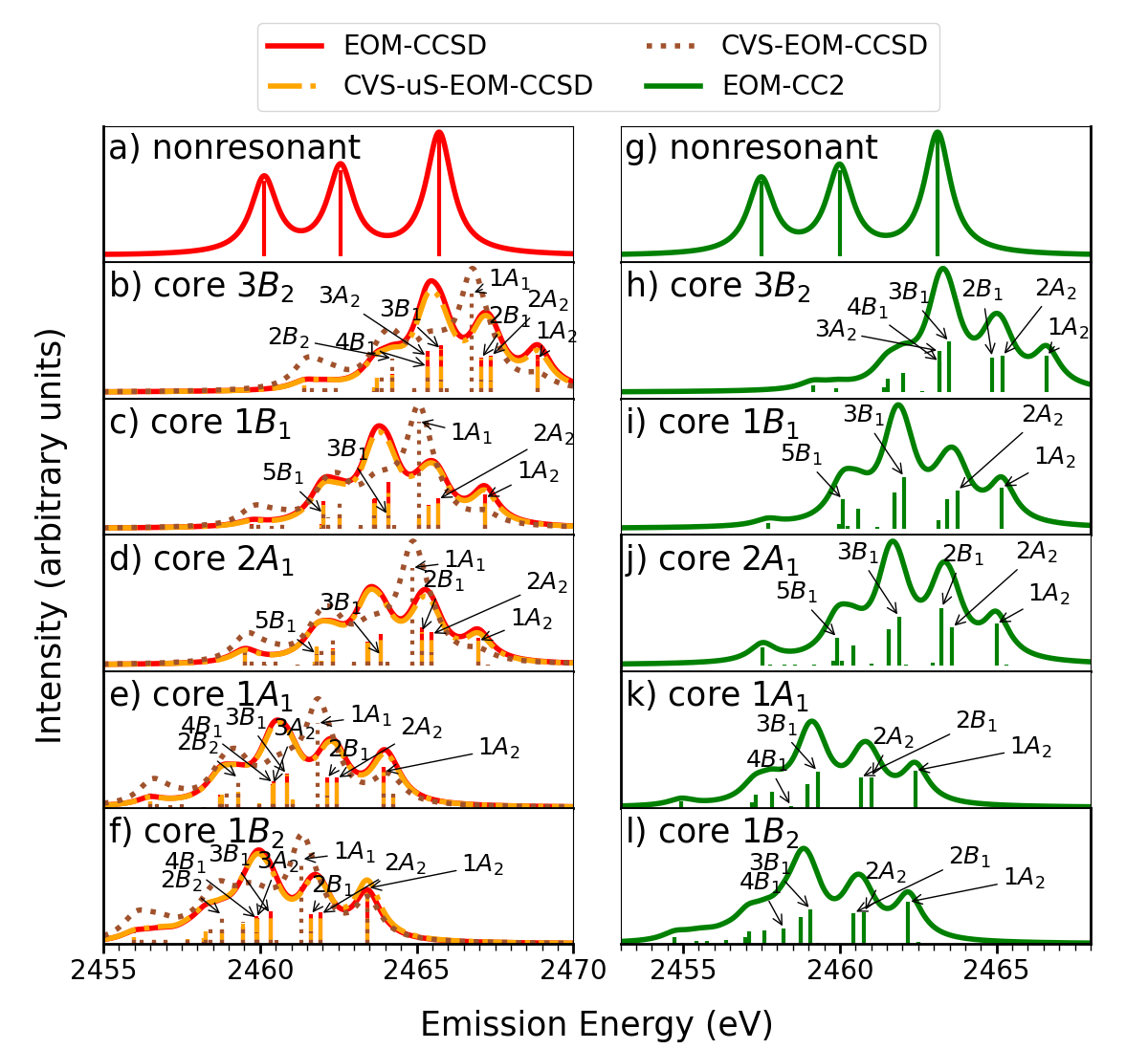}%{Figures/new_RIXS_h2s.png}
    \caption{H$_2$S: Nonresonant (a and g) and RIXS spectra at resonance with the energy of the 1$B_2$ core excitation (f and l), 1$A_1$ core excitation (e and k), 2$A_1$ core excitation (d and j), 2$B_2$ core excitation (c and i) and 3$B_2$ core excitation (b and h) at the S $K$-edge. All spectra are computed with the 6-311++G**. No shift has been applied to the results. A Lorentzian broadening was applied with HWHM=0.51eV. Arrows
indicate the main transitions probed. Note that $1A_1$ labels the first valence excitation of $A_1$ symmetry.}
    \label{fig:h2s_rixs_all}
\end{figure}
\begin{table}[h]
    \caption{H$_2$S: Overview of valence states probed in RIXS by the different methods at the considered core resonances corresponding to the low energy XAS peak. As CVS-uS-EOM-CCSD and EOM-CCSD results appear identical, these are only reported once under EOM-CCSD. Energies reported here are not shifted.  The label $1A_1$ denotes the first valence excitation of $A_1$ symmetry.
    }
    \label{tab:h2s_rixs_assign_res1}
    \centering
    \begin{tabular}{cccccc}
    \hline
    \multicolumn{2}{c}{EOM-CCSD}&\multicolumn{2}{c}{CVS-EOM-CCSD}&\multicolumn{2}{c}{EOM-CC2}\\
         Energy/ eV& Probed state & Energy/ eV& Probed state  & Energy/ eV& Probed state\\\hline
         \multicolumn{6}{c}{core 1$B_2$ resonance}\\\hline
         2463.4265&1$A_2$&2463.4265&1$A_2$&2462.1849&1$A_2$\\
         %2461.9345& 2$A_2$ &2461.9345& -& 2460.7734& 2$A_2$\\
         2461.9345& 2$A_2$ &-& -& 2460.7734& 2$A_2$\\
         %2461.6200&2$B_1$&2461.6200&-&2460.4409&2$B_1$\\
         2461.6200&2$B_1$&-&-&2460.4409&2$B_1$\\
         %2461.3149&-&2461.3149&1$A_1$&2460.1634 &-\\
         -&-&2461.3149&1$A_1$&- &-\\
         %2460.3345 & 3$B_1$&2460.3345 & - &2459.0692 &3$B_1$\\
         2460.3345 & 3$B_1$&- & - &2459.0692 &3$B_1$\\
         %2459.9138 & 4$B_1$ &2459.9138 & - &2458.7552&3$A_2$\\
         2459.9138 & 4$B_1$ &- & - &2458.7552&3$A_2$\\
         %2459.8942 & 3$A_2$& 2459.8942 & - & 2458.7546&4$B_1$\\
         2459.8942 & 3$A_2$& - & - & 2458.7546&4$B_1$\\
         %2458.7911 &-&2458.7911 & 2$B_2$&2457.6041&-
         - &-&2458.7911 & 2$B_2$&-&-
         \\\hline
             \multicolumn{6}{c}{core 1$A_1$ resonance}\\\hline
         2463.9417&1$A_2$&2463.9417&1$A_2$&2462.4078&1$A_2$\\
         %2462.4494& 2$A_2$ &2462.4494& -& 2460.9963& 2$A_2$\\
         2462.4494& 2$A_2$ &-& -& 2460.9963& 2$A_2$\\
         %2462.1351&2$B_1$&2462.1351&-&2460.6638&2$B_1$\\
         2462.1351&2$B_1$&-&-&2460.6638&2$B_1$\\
         %2461.8298&-&2461.8298&1$A_1$&2460.3862 &-\\
         -&-&2461.8298&1$A_1$&- &-\\
         %2460.8494  & 3$B_1$&2460.8494  & - &2459.292 &3$B_1$\\
         2460.8494  & 3$B_1$&-  & - &2459.292 &3$B_1$\\
         %2460.4289 & 4$B_1$ &2460.4289 & - &2458.9780&3$A_2$\\
         2460.4289 & 4$B_1$ &- & - &2458.9780&3$A_2$\\
         %2460.4093 & 3$A_2$& 2460.4093 & - & 2458.9775&4$B_1$\\
         2460.4093 & 3$A_2$& - & - & 2458.9775&4$B_1$\\
         %2459.3059 &-&2459.3059 & 2$B_2$&2457.8273&-
         - &-&2459.3059 & 2$B_2$&-&-
         \\\hline
    \end{tabular}
\end{table}
\begin{table}[h]
    \caption{H$_2$S: Overview of valence states probed in RIXS by the different methods at the considered core resonances corresponding to the middle peak of the XAS. As CVS-uS-EOM-CCSD and EOM-CCSD results appear identical, these are only reported once under EOM-CCSD. Energies reported here are not shifted.  The label $1A_1$ denotes the first valence excitation of $A_1$ symmetry.
    }
    \label{tab:h2s_rixs_assign_res2}
    \centering
    \begin{tabular}{cccccc}
    \hline
    \multicolumn{2}{c}{EOM-CCSD}&\multicolumn{2}{c}{CVS-EOM-CCSD}&\multicolumn{2}{c}{EOM-CC2}\\
         Energy/ eV& Probed state & Energy/ eV& Probed state  & Energy/ eV& Probed state\\\hline
         \multicolumn{6}{c}{core 2$A_1$ resonance}\\
         \hline
         %2466.9583&1$A_2$&2466.9583&-&2464.9966&1$A_2$\\
         2466.9583&1$A_2$&-&-&2464.9966&1$A_2$\\
         %2465.4663& 2$A_2$ &2465.4663& -& 2463.5852 & 2$A_2$\\
         2465.4663& 2$A_2$ &-& -& 2463.5852 & 2$A_2$\\
         %2465.1517&2$B_1$&2465.1517&-&2463.2529&2$B_1$\\
         2465.1517&2$B_1$&-&-&2463.2529&2$B_1$\\
         %2464.8464&-&2464.8464&1$A_1$&2462.9754&-\\
        -&-&2464.8464&1$A_1$&-&-\\
         %2463.8660 & 3$B_1$&2463.8660 &-&2461.8812 &3$B_1$\\
         2463.8660 & 3$B_1$&- &-&2461.8812 &3$B_1$\\
         %2461.8064& 5$B_1$ &2461.8064&-&2459.9136&5$B_1$
         2461.8064& 5$B_1$ &-&-&2459.9136&5$B_1$
         \\\hline
         \multicolumn{6}{c}{core 1$B_1$ resonance}\\\hline
         %2467.1733&1$A_2$&2467.1733&-&2465.1694&1$A_2$\\
         2467.1733&1$A_2$&-&-&2465.1694&1$A_2$\\
         %2465.6813& 2$A_2$ &2465.6813& -& 2463.7580& 2$A_2$\\
         2465.6813& 2$A_2$ &-& -& 2463.7580& 2$A_2$\\
         %2465.0617&-&2465.0617&1$A_1$&2463.1482&-\\
         -&-&2465.0617&1$A_1$&-&-\\
         %2464.0810 & 3$B_1$&2464.0810 &-&2462.0540 &3$B_1$\\
         2464.0810 & 3$B_1$&- &-&2462.0540 &3$B_1$\\
         %2462.0216& 5$B_1$ &2462.0216&-&2460.0863&5$B_1$
         2462.0216& 5$B_1$ &-&-&2460.0863&5$B_1$
         \\\hline
    \end{tabular}
\end{table}
\begin{table}[h]
%\vspace{-1.5cm}
    \caption{H$_2$S: Overview of valence states probed in RIXS by the different methods at the considered core resonance corresponding to the small XAS peak found just below the ionization threshold. As CVS-uS-EOM-CCSD and EOM-CCSD results appear identical, these are only reported once under EOM-CCSD. Energies reported here are not shifted.  The label $1A_1$ denotes the first valence excitation of $A_1$ symmetry.
    }
    \label{tab:h2s_rixs_assign_res3}
    \centering
    \begin{tabular}{cccccc}
             \hline
    \multicolumn{2}{c}{EOM-CCSD}&\multicolumn{2}{c}{CVS-EOM-CCSD}&\multicolumn{2}{c}{EOM-CC2}\\
         Energy/ eV& Probed state & Energy/ eV& Probed state& Energy/ eV& Probed state  \\\hline
         \multicolumn{6}{c}{core 3$B_2$ resonance}\\\hline
         %2468.8686&1$A_2$&2468.8686&-&2466.6043&1$A_2$\\
         2468.8686&1$A_2$&-&-&2466.6043&1$A_2$\\
         %2467.3766&2$A_2$&2467.3766&-&2465.1928&2$A_2$\\
         2467.3766&2$A_2$&-&-&2465.1928&2$A_2$\\
         %2467.0620&2$B_1$&2467.0620&-&2464.8603&2$B_1$\\
         2467.0620&2$B_1$&-&-&2464.8603&2$B_1$\\
         %2466.7569&-&2466.7569&1$A_1$&2464.5828&-\\
         -&-&2466.7569&1$A_1$&-&-\\
         %2465.7765 &3$B_1$&2465.7765 &-&2463.4886&3$B_1$\\
         2465.7765 &3$B_1$&- &-&2463.4886&3$B_1$\\
         %2465.3558 &-& 2465.3558 &-&2463.1756&3$A_2$\\
         2465.3558 &$4B_1$& - &-&2463.1756&3$A_2$\\
         %2465.3362 &3$A_2$& 2465.3362 &-&2463.1740&4$B_1$\\
         2465.3362 &3$A_2$& - &-&2463.1740&4$B_1$\\
         %2464.2331&-&2464.2331&2$B_2$&2462.0235&-
         -&-&2464.2331&2$B_2$&-&-
         \\\hline
    \end{tabular}
\end{table}

Note that the RIXS cross sections are approximately one order of magnitude lower than those computed for H$_2$O and CH$_3$OH (see Sections \ref{SI-sec:h2o}-\ref{SI-sec:h2s} of the SI). This is also the case for the XAS oscillator strengths.%Inspecting the results collected in Tables \ref{SI-tab:h2s_rixs_1st}-\ref{SI-tab:h2s_xes_S} in the SI (Section \ref{SI-sec:h2s}), it is seen that the RIXS cross sections are approximately one order of magnitude lower than those computed for H$_2$O and CH$_3$OH (Sections \ref{SI-sec:h2o} and \ref{SI-sec:ch3oh}). This is also the case for the XAS oscillator strengths.

Considering the RIXS spectra in Figs.~\ref{fig:h2s_rixs_all}b-f, it is evident that employing the ``full'' CVS approximation when solving the damped response equations significantly affects the spectra. While we observe a good correspondence between the (unprojected) EOM-CCSD results and the CVS-uS-EOM-CCSD results, as well as the EOM-CC2 results in Figs. \ref{fig:h2s_rixs_all}h-l (disregarding the energy shift), a different spectral shape is obtained when employing the full CVS approximation. This was also observed by \citeauthor{2020_Nanda_JCP} for the molecule para-nitro-aniline.\cite{2020_Nanda_JCP} 

The shifts between the CC2 and CCSD results are found to change at higher resonances. Thus, at the first two resonances the shift is approximately 1 eV (Figs. \ref{fig:h2s_rixs_all}e, \ref{fig:h2s_rixs_all}f, \ref{fig:h2s_rixs_all}k and \ref{fig:h2s_rixs_all}l), while it is approximately 2.5 eV at the third and fourth resonances (Figs.  \ref{fig:h2s_rixs_all}c, \ref{fig:h2s_rixs_all}d, \ref{fig:h2s_rixs_all}i and \ref{fig:h2s_rixs_all}j) and approximately 2 eV at the last considered resonance (Figs. \ref{fig:h2s_rixs_all} b and h). The shift between the nonresonant CCSD and CC2 spectra is approximately 3 eV (Figs. \ref{fig:h2s_rixs_all}a and \ref{fig:h2s_rixs_all}g). This is in line with our observations for CH$_3$OH, where the shift applied at the first resonance to align with experiment (Figs. \ref{fig:methanol_rixs_O}c and \ref{fig:methanol_rixs_O}f) was noted to no longer be sufficient at higher resonances, but to different degrees for CCSD and CC2 (Figs. \ref{fig:methanol_rixs_O}a,b and \ref{fig:methanol_rixs_O}d,e). As for CH$_3$OH, this observation is attributed to differences in the XAS, and for the nonresonant case, differences in the core and valence IEs.

At the first two resonances 
(Figs.~\ref{fig:h2s_rixs_all}e, \ref{fig:h2s_rixs_all}f, \ref{fig:h2s_rixs_all}k and \ref{fig:h2s_rixs_all}l) the valence states probed are $1A_2$ as well as $2A_2$, $2B_1$, $3B_1$, $3A_2$, and $4B_1$ for both EOM-CCSD, EOM-CVS-uS-CCSD and EOM-CC2 methods (see Table \ref{tab:h2s_rixs_assign_res1}). Likewise, these methods predict that at the next two resonances (Figs. \ref{fig:h2s_rixs_all}c, \ref{fig:h2s_rixs_all}d, \ref{fig:h2s_rixs_all}i and \ref{fig:h2s_rixs_all}j), associated with the second peak in the XAS, the valence states probed are the same except for $4B_1$ and $3A_2$, which are replaced by $5B_1$ (see Table \ref{tab:h2s_rixs_assign_res2}). Finally, EOM-CCSD and EOM-CVS-uS-CCSD predict once more the same valence transitions to be probed at the core 3$B_2$ resonance (Figs. \ref{fig:h2s_rixs_all}b and \ref{fig:h2s_rixs_all}h) as at the first two resonances (see Table \ref{tab:h2s_rixs_assign_res3}). The EOM-CVS-CCSD calculation indicates that it is mainly $1A_1$ that is probed at all resonances. In addition, at this level of theory, the $2B_2$ state also appears to be probed at the first, second, and fifth resonances (see Tables \ref{tab:h2s_rixs_assign_res1}-\ref{tab:h2s_rixs_assign_res3}). %Likewise, the $2B_2$ state is probed at the third and fourth resonances, while the $4B_1$, $2B_2$ and $3A_1$ states are probed at the last considered resonance.
In general, we observe that for this system it is the same states that are probed at all resonances unlike what was seen for H$_2$O. Note that here, transitions of the same symmetry for the different methods also display the same characters between the methods (see Fig. \ref{SI-fig:H2S_NTOs}).

The overall spectral shape showing three main features with varying relative intensities at all resonances is in agreement with experiment.\cite{Kavcic_2009}
\FloatBarrier
\subsubsection{Para-nitro-aniline (PNA)}
For the larger molecule PNA, both the XAS and valence absorption spectra show small discrepancies between CCSD and CC2 (see Figure \ref{SI-fig:pna_xas}).
However, both methods predict the first bright core transition to be core $1B_1$.

%
%\revS{IS THERE A PROBLEM WITH THIS TABLE? uS and full CVS numbers are identical}
%NO the energies are the same for CVS and CVS-uS, only intensities might change (so the probed state column)
\setlength{\tabcolsep}{6pt}
 \begin{table}[h]
 \small
    \centering
        \caption{PNA: Overview of valence states probed in RIXS with the different methods at the first core resonance. All calculations employed the 6-311++G** basis set. Energies reported here are not shifted. The label $1A_1$ denotes the first valence excitation of $A_1$ symmetry.
        }
    \label{tab:pna_rixs_assign_res1}
    \begin{tabular}{cccccccc}
    \hline
    \multicolumn{2}{c}{CVS-uS-EOM-CCSD}&\multicolumn{2}{c}{CVS-EOM-CCSD}&\multicolumn{2}{c}{fc-CVS-0-EOM-CCSD}&\multicolumn{2}{c}{CVS-uS-EOM-CC2}\\
         Energy& Probed & Energy& Probed  & Energy& Probed & Energy& Probed \\
       / eV  & state & / eV & state & / eV & state  &
       / eV & state\\
         \hline
         \multicolumn{8}{c}{core 1$B_1$ resonance}\\\hline
       282.0633&1$A_1$&282.0633&1$A_1$&281.5675&1$A_1$&283.5360&1$A_1$\\
       280.3044&2$A_1$&280.3044& 2$A_1$&279.8086&2$A_1$&281.7447&2$A_1$\\
       279.5582&3$A_1$&279.5582&3$A_1$&279.0722&3$A_1$& 280.9585&3$A_1$\\
       278.9345&5$A_1$&278.9345&5$A_1$&278.4407&5$A_1$& -&-
         \\\hline
    \end{tabular}
\end{table}
\setlength{\tabcolsep}{10pt}

\begin{figure}
    \centering
    \includegraphics[width=\textwidth]{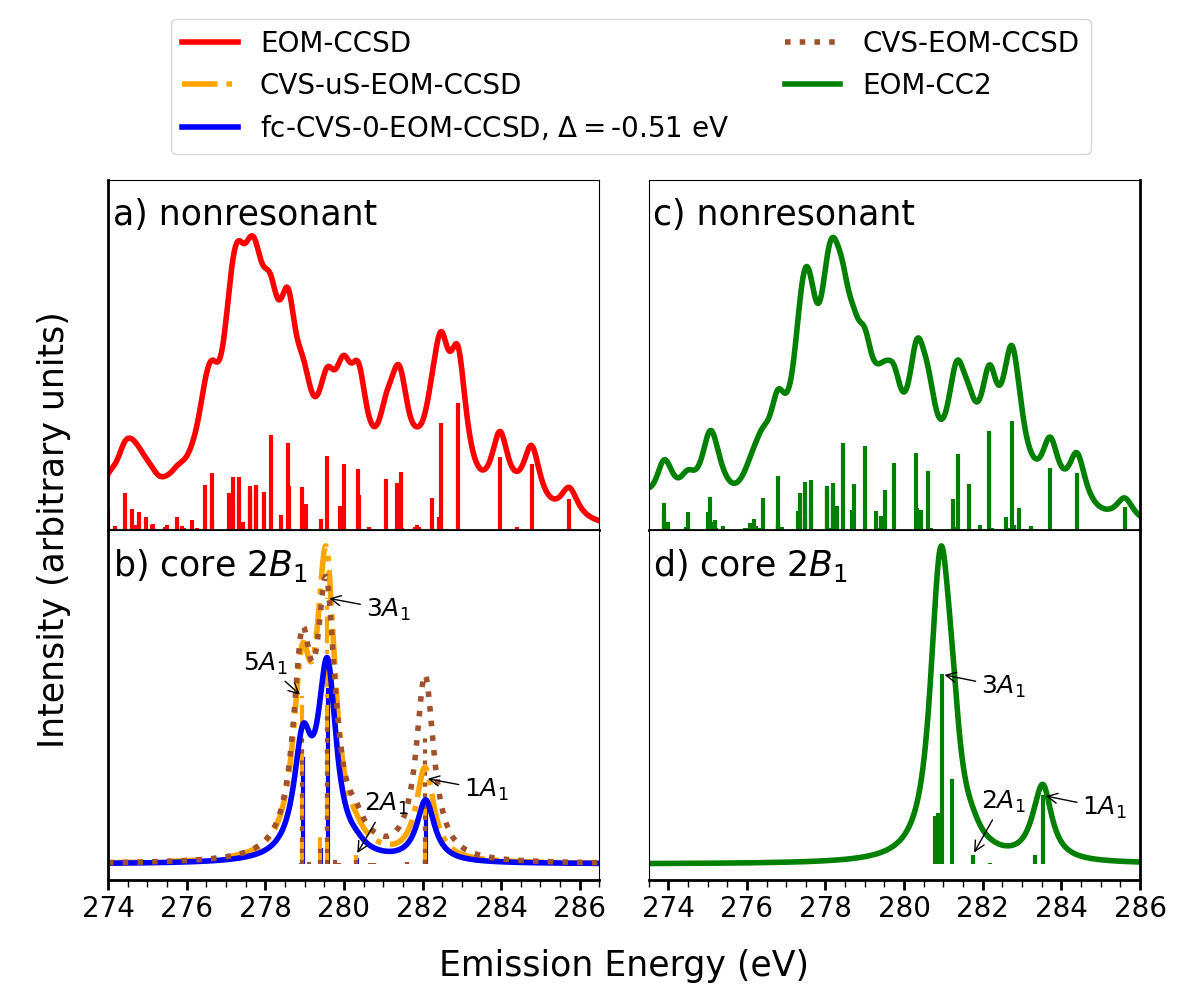}%{Figures/new_rixs_pna_C.png}
    \caption{PNA: Nonresonant (a and c) and RIXS spectra at resonance with the energy of the first core excitation (b and d) at the C $K$-edge. The 6-311++G** basis set was employed in all calculations. The fc-CVS-0-EOM-CCSD results were shifted to align with the EOM-CVS-uS-CCSD results. A Lorentzian broadening was applied with HWHM=0.27 eV. Arrows
indicate the main transitions probed. Note that $1A_1$ labels the first valence excitation of $A_1$ symmetry.}
    \label{fig:pna_rixs_all}
\end{figure}

The nonresonant emission spectra, as well as the RIXS spectra resonant with the first bright core excitation 
at the C $K$-edge (computed at 288.0 eV at the EOM-CC2 level and at 
286.7 eV at the EOM-CCSD level of theory without applying a shift), are shown in Fig. \ref{fig:pna_rixs_all}. In addition, the RIXS spectrum has been computed at the fc-CVS-0-EOM-CCSD level of theory (core $1B_1$ resonance at 288.9 eV). Here, we observe that the RIXS spectra at the CCSD level look similar, except for the intensity of the high energy feature. This feature is much more intense, when the full CVS approximation is employed only for solving the divergent damped response equations (labeled CVS-EOM-CCSD). In the fc-CVS-0-EOM-CCSD calculation (full CVS used when solving \textit{all} damped response equations), this is not the case. It was found (see SI, Section \ref{SI-sec:pna}) that it is indeed the difference between CVS and CVS-0 rather than the use of fc that gives rise to the large difference.

The same three valence states are probed in all CCSD based calculations, namely $1A_1$, $3A_1$ and $5A_1$ (see Table \ref{tab:pna_rixs_assign_res1}). The CC2 spectrum looks similar to the CCSD ones, but it is observed that the low energy transitions are much more closely spaced, and thus we observe only two peaks rather than three. The transitions probed at the CC2 level are again $1A_1$ and $3A_1$ (see Table \ref{tab:pna_rixs_assign_res1}). We note that not only the symmetries, but also the electronic characters of these transitions are the same at the CC2 and CCSD levels of theory (see Fig. \ref{SI-fig:pna_NTOs}). As for the nonresonant spectra, the CC2 and CCSD spectra are somewhat similar in shape, with an additional shift of the CC2 spectrum compared to CCSD, as also observed for the other systems.

Our CVS-uS-EOM CCSD results are in good agreement with those presented by \citeauthor{2020_Nanda_JCP}.\cite{2020_Nanda_JCP} It is, however, noted that our CVS-EOM-CCSD and CVS-uS-EOM-CCSD spectra are rather similar to each other. This is in contrast to the results presented by \citeauthor{2020_Nanda_JCP},\cite{2020_Nanda_JCP} where large discrepancies between CVS and CVS-uS based results were reported. Nonetheless, our calculation utilizing the fc-CVS-0-EOM-CCSD level of theory, showed good agreement with our other results (see Fig. \ref{fig:pna_rixs_all}b,d and Table \ref{SI-tab:pna_rixs_1st} in Section \ref{SI-sec:pna}). Discrepancies were, as mentioned, attributed to the fact that we employ the CVS approximation only when solving amplitude response equations for positive frequencies and multiplier response equations for negative frequencies. The implementation by Nanda et al.,\cite{2020_pccp_Nanda_RIXS, 2020_Nanda_JCP} on the other hand, employs CVS when solving all damped response equations.
\FloatBarrier
\subsection{Evaluation of performance of CCSD for a solvated molecule}\label{sec:imidazole}
\begin{figure}
    \centering
    \includegraphics[width=0.4\textwidth]{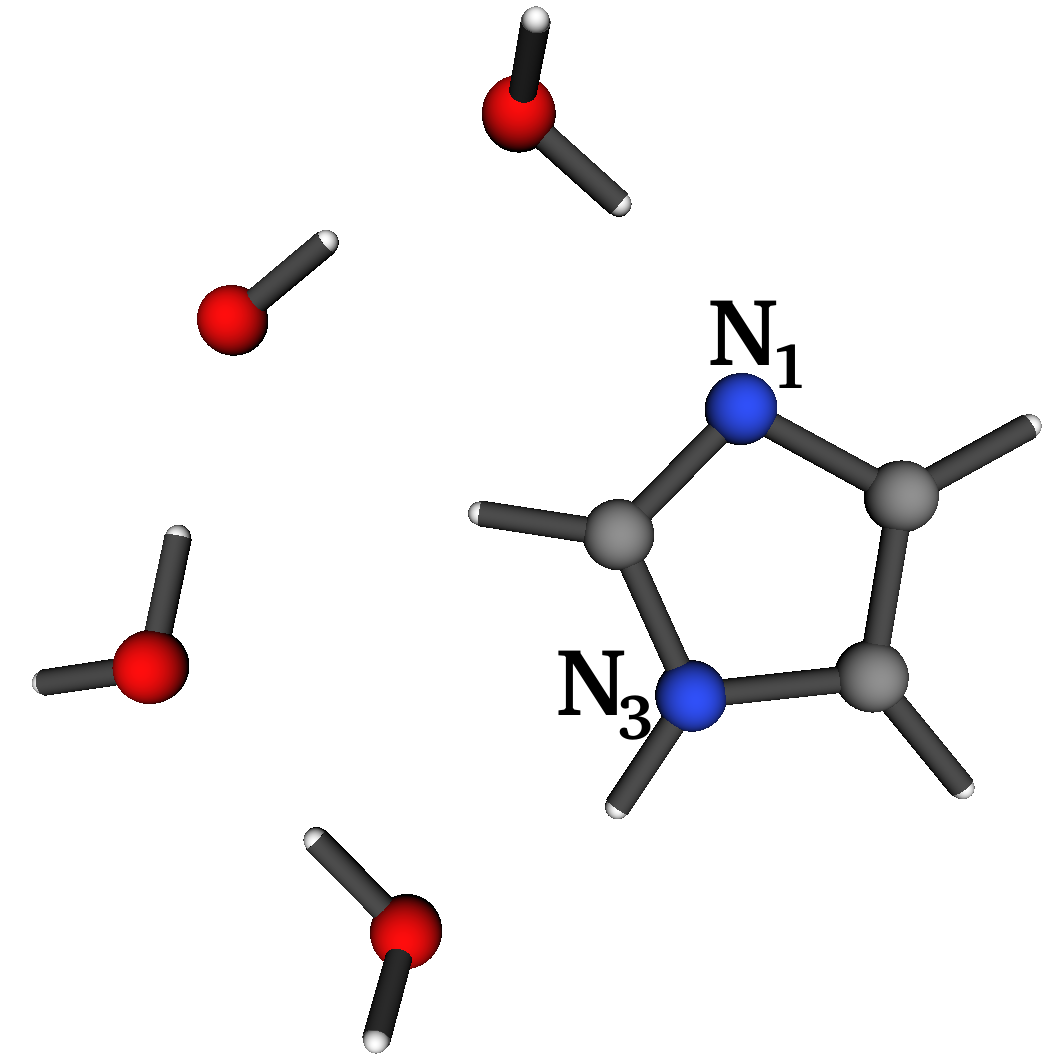}%{Figures/imi_4whf2.png.png}
    \caption{Structure of imidazole with 4 explicit H$_2$O molecules.}
    \label{fig:imi4w_struct}
\end{figure}

X-ray experiments have been carried out in aqueous solution for imidazole.\cite{meyer_18} Previous studies have shown that including explicit H$_2$O molecules can yield good results,\cite{jagoda_8, Thomasonphd_12, gougoula_20} although, one H$_2$O molecule is not enough for describing the XAS spectrum.\cite{Thomasonphd_12} For the best results, 15 or more water molecules should be included.\cite{Thomasonphd_12} We have chosen to include only 4 explicit H$_2$O molecules (see structure in 
Fig. \ref{fig:imi4w_struct}), as this was shown by \citeauthor{Thomasonphd_12}\cite{Thomasonphd_12} to yield reasonable results and is computationally feasible even when treating the entire system at the CCSD level of theory.
\FloatBarrier
To reduce the computational cost in the RIXS calculations, we treated the solvent molecules at the Hartree-Fock (HF) level of theory and only employed coupled cluster methods for the imidazole molecule itself. In this ``CCSD-in-HF'' approach, HF orbitals localized on the region of interest (in our case, imidazole) are included in the CCSD calculation. The remaining orbitals enter the coupled cluster equations through their contribution to the Fock matrix.\citep{sanchez2010cholesky} 
In this work, we use Cholesky orbitals\citep{aquilante2006fast,sanchez2010cholesky} for the occupied space and orthonormalized projected atomic orbitals\citep{pulay1983,saebo1993} for the virtual space. The CCSD-in-HF approach has been demonstrated to work well for localized transitions\citep{folkestad2020equation} such as those addressed in XAS. 
However, its accuracy relies on a sufficiently large active region to describe the transitions of interest, i.e., delocalized excitations that go outside of the active region cannot be described.

The simulated XAS was found to correspond well with experiment as well as with the calculation where the H$_2$O molecules are also treated at the CC level of theory (see Fig. \ref{SI-fig:xas_imiw4_N}, Section \ref{SI-sec:imi}). Likewise, the simulated valence absorption spectrum (also in Fig. \ref{SI-fig:xas_imiw4_N}) was found to change only little when treating the solvent molecules at the HF level of theory compared to treating the entire system at the CC level of theory.
We thus concluded that no significant accuracy is lost by treating the solvent molecules at a lower level of theory. This was expected for XAS, since the core transitions are very localized.

\begin{figure}
    \centering
    \includegraphics[width=\textwidth]{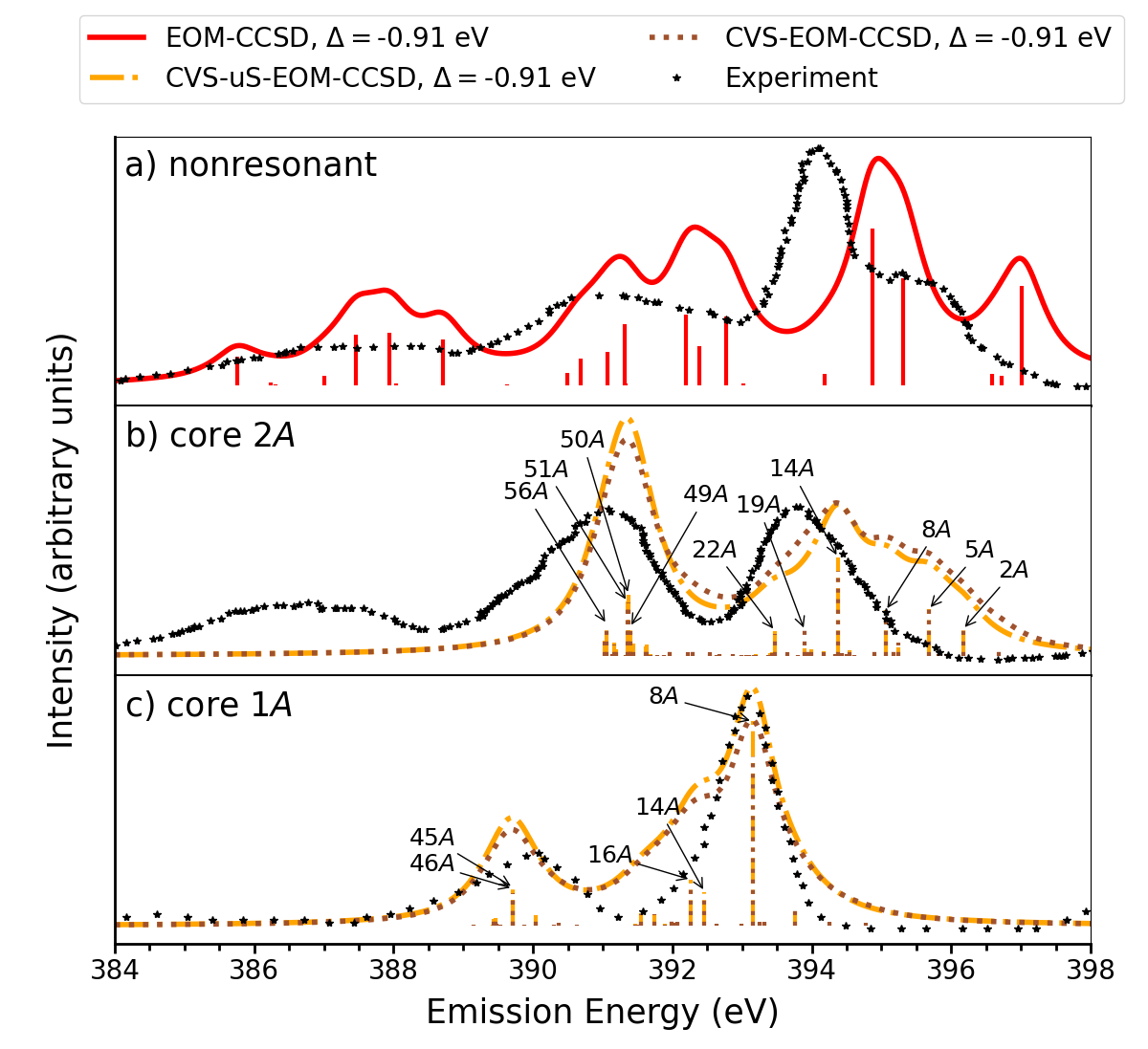}%{Figures/new_rixs_imi4whf_ccsd.png}
    \caption{Imidazole with 4 explicit H$_2$O molecules,
    the latter treated at the HF level of theory: Nonresonant (a) and RIXS spectra at resonance with the energy of the first (c), and second (b) core excitation at the N $K$-edge. The results are shifted by the amounts indicated in the legend to align to high energy peak of the RIXS experiment at resonance with the first core excitation. Experimental data in aqueous solution was digitized from \citeauthor{meyer_18}\cite{meyer_18} A Lorentzian broadening has been applied with HWHM=0.41 eV. Arrows indicate the main valence transitions probed. Note that $1A$ denotes the first valence excitated state.}
    \label{fig:imi4whf_rixs_all}
\end{figure}

\begin{figure}
     \centering
     \includegraphics[width=\textwidth]{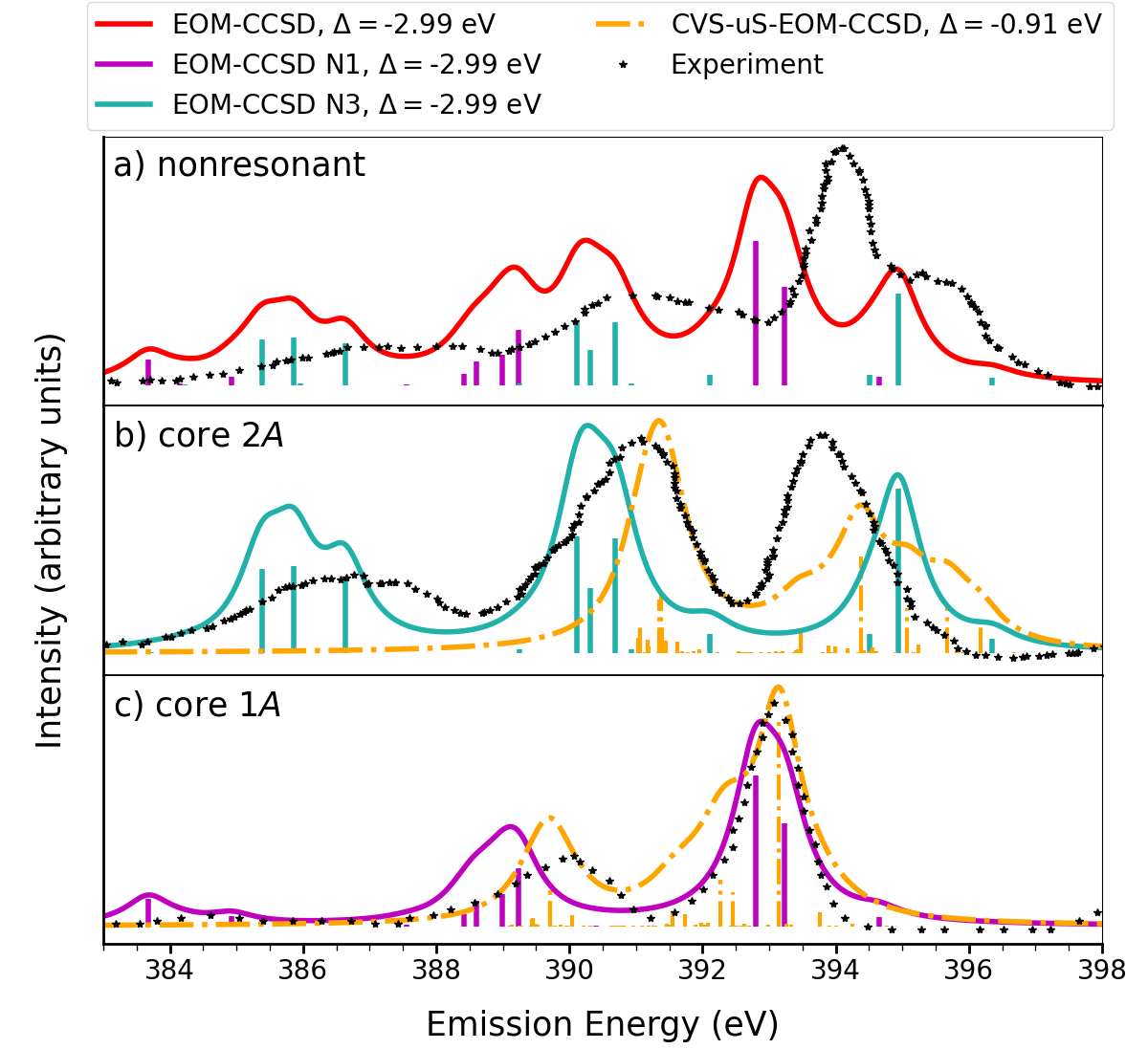}%{Figures/new_rixs_imi4whf_nonres_ccsd.png}
     \caption{Imidazole with 4 explicit H$_2$O molecules, the latter treated at the HF level of theory: Nonresonant XES (a) and RIXS spectra at resonance with the energy of the first (c), and second (b) core excitation at the N $K$-edge. The results are shifted by the amounts indicated in the legends to aling to the high energy peak of the RIXS experiment\cite{meyer_18} at resonance with the first core excitation. 
     All calculations employed the 6-311++G** basis set and a Lorentzian broadening with HWHM=0.41 eV was applied. For (b) and (c) the simulated spectra are the nonresonant ones of individual N as indicated in the legend. The CVS-uS-EOM-CCSD RIXS calculations are also shown. Experimental data in aqueous solution was digitized from \citeauthor{meyer_18}\cite{meyer_18}}
     \label{fig:imi4whf_rix_nonres}
 \end{figure}

We computed the RIXS spectra at resonance with the first two intense core transitions in the XAS spectrum, labeled ``core $1A$'' and 
``core $2A$''.
Note that the ``core $1A$'' resonance is primarily 
due to the excitation of a $1s$(N1) electron, whereas the ``core $2A$'' resonance is dominated by the excitation of an electron in the $1s$ orbital of N3. 
 The nonresonant XES and resonant RIXS spectra can be found in Fig.~\ref{fig:imi4whf_rixs_all}, and the valence states probed by RIXS are found in Table \ref{tab:imihf_rixs_assign_res}. 
\begin{table}[h]
    \centering
    \caption{Imidazole with 4 explicit H$_2$O molecules, the latter treated at the HF level of theory: Overview of valence states probed in RIXS at the first and second core resonance. All calculations employed the 6-311++G** basis set. Tabulated energies have not been shifted. The label $1A$ denotes the first valence excitation.
        }
    \label{tab:imihf_rixs_assign_res}
    \begin{tabular}{cccc}
    \hline
    \multicolumn{2}{c}{CVS-uS-EOM-CCSD}&\multicolumn{2}{c}{CVS-EOM-CCSD}\\
         Energy/ eV& Probed state & Energy/ eV& Probed state  \\\hline
         \multicolumn{4}{c}{core 1$A$ resonance}\\
         \hline
 394.0603 &  8A & 394.0603 &  8A \\
 393.3704 & 14A & 393.3704 & 14A \\
 393.1745 & 16A & 393.1745 & 16A \\
 390.6254 & 45A& 390.6254  & 45A \\
 390.6188 & 46A & 390.6188 & 46A \\
         \hline
         \multicolumn{4}{c}{core 2$A$
         resonance}\\\hline
 397.0810 & 2A & 397.0810 & 2A \\
 396.5893 & 5A & 396.5893 & 5A \\
 395.9735 & 8A & 395.9735 & 8A \\
 395.2837 & 14A &395.2837 & 14A \\
 %394.8075 & - & 394.8075 & 19A \\
  - & - & 394.8075 & 19A \\
 394.3794 & 22A & 394.3794 & 22A \\
 392.3035 & 49A & 392.3035 & 49A \\
 392.2825 & 50A & 392.2825 & 50A \\
392.2616 & 51A & 392.2616 & 51A \\
391.9669 & 56A & 391.9669 & 56A \\\hline
    \end{tabular}
\end{table}
\citeauthor{meyer_18}\cite{meyer_18} analyzed the experimental RIXS spectra at the two  mentioned resonances by considering only computed nonresonant emission spectra for each individual nitrogen. This has also been done here and can be seen in 
Fig.~\ref{fig:imi4whf_rix_nonres}. 
 In line with the localized nature (on specific N atoms) of the resonant core transitions considered, the nonresonant emission spectrum of N1 reproduces the spectral shape of the first resonance (see Fig.~\ref{fig:imi4whf_rix_nonres}c). Likewise, the spectral shape of the RIXS at the second core resonance is reasonably well captured by the nonresonant emission spectrum of N3 (see Fig. \ref{fig:imi4whf_rix_nonres}b), though not to the same degree as seen for the first resonance. 
 
 As discussed by \citeauthor{meyer_18}\cite{meyer_18},
 this XES-based
 analysis of the RIXS 
  spectral slices highlights the 
  fact that spectral changes are mainly due to the selective excitation
  of one particular nitrogen site. This approach is also computationally much cheaper than the 
  computation of the full RIXS cross section.

 Calculating the RIXS spectra as we do here by using the damped response formalism requires solving the eigenvalue problem for a large number of roots. Moreover, a correspondingly large number of damped equations must be solved to obtain the response amplitudes and multipliers that enter the RIXS transition moments, which is computationally expensive and may be difficult to converge. On the other hand, it allows for an assignment of the spectral bands in terms of probed valence excited states (as we probe the correlation between valence and core excitations). Furthermore, when considering the second resonance (Figs. \ref{fig:imi4whf_rixs_all}b and \ref{fig:imi4whf_rix_nonres}b), the separation between the two peaks is 
 better reproduced by the RIXS calculation. The latter, however, does not capture the low energy peak of the spectrum with the considered number of roots. This peak on the other hand, is captured when simply considering the nonresonant calculation even for a limited number of roots. In general, our results show that, while the calculation is computationally challenging, the experimental results are well reproduced also for the solvated molecule.
 Other molecules in solution should also be considered, to verify whether this approach is capable of yielding good results for RIXS of solvated molecules in general. Also, for a better description of experiment, more than one molecular structure should be considered. This, however, is beyond the scope of this study.
 
\section{Conclusions}
A new implementation of a damped response solver for the calculation of, amongst others, RIXS spectra at the EOM-CCSD and EOM-CC2 levels of theory in $e^T$ was presented. It was found that while CVS-EOM-CC2 generally does not yield XAS spectra in sufficiently good agreement with experiment compared to CVS-EOM-CCSD, the RIXS spectra are often similar at the two levels of theory, except for an overall energy shift. For the second core resonance of 
%H$_2$O and 
CH$_3$OH, as well as the first core resonance of PNA, discrepancies were observed at the edge of the explored energy region. This, however, might be remedied by considering a larger number of roots. In addition, the valence excitations probed are in overall good agreement between methods. Some discrepancies can, however, be found and care must be taken when assigning the states. Still, the EOM-CC2 results are promising and the method can become a valuable tool for performing RIXS calculations.
Further studies are, however, needed to assess the full capabilities of the method compared to EOM-CCSD. 
Furthermore, it was found that the CVS-uS model proposed by \citeauthor{2020_Nanda_JCP} might improve RIXS results significantly compared to the regular CVS model even for smaller systems, as exemplified by H$_2$S. Finally, our EOM-CCSD implementation of RIXS was tested on imidazole in an aqueous solution. Even though more studies are required to explore its capabilities in this respect, the results are, also in this case, quite promising.

\begin{acknowledgement}
AKSP acknowledges support from the DTU Partnership PhD programme. SDF acknowledges funding from the Research Council of Norway through FRINATEK project 275506.
SC acknowledges support from 
the Independent Research Fund Denmark (DFF-RP2 Grant 7014-00258B).
The European Cooperation in Science and Technology, COST Action CA18222 {\it Attochem} is also acknowledged.
\end{acknowledgement}
%%%%%%%%%%%%%%%%%%%%%%%%%%%%%%%%%%%%%%%%%%%%%%%%%%%%%%%%%%%%%%%%%%%%%
\begin{suppinfo}

The Supporting Information contains:  Geometries of the investigated molecules;
the 6-311++G** basis set including additional Rydberg functions used; all data tables for the plotted spectra;  XAS and valence absorption spectra for CH$_3$OH, H$_2$S, PNA, and imidazole with 4 explicit water molecules; comparisons of fc-CVS-0-EOM-CCSD, CVS-0-EOM-CCSD and CVS-EOM-CCSD results for PNA using a small basis; results of a calculation on pyridine.

\end{suppinfo}
\begin{tocentry}

\includegraphics[width=3.25in,height=1.75in]{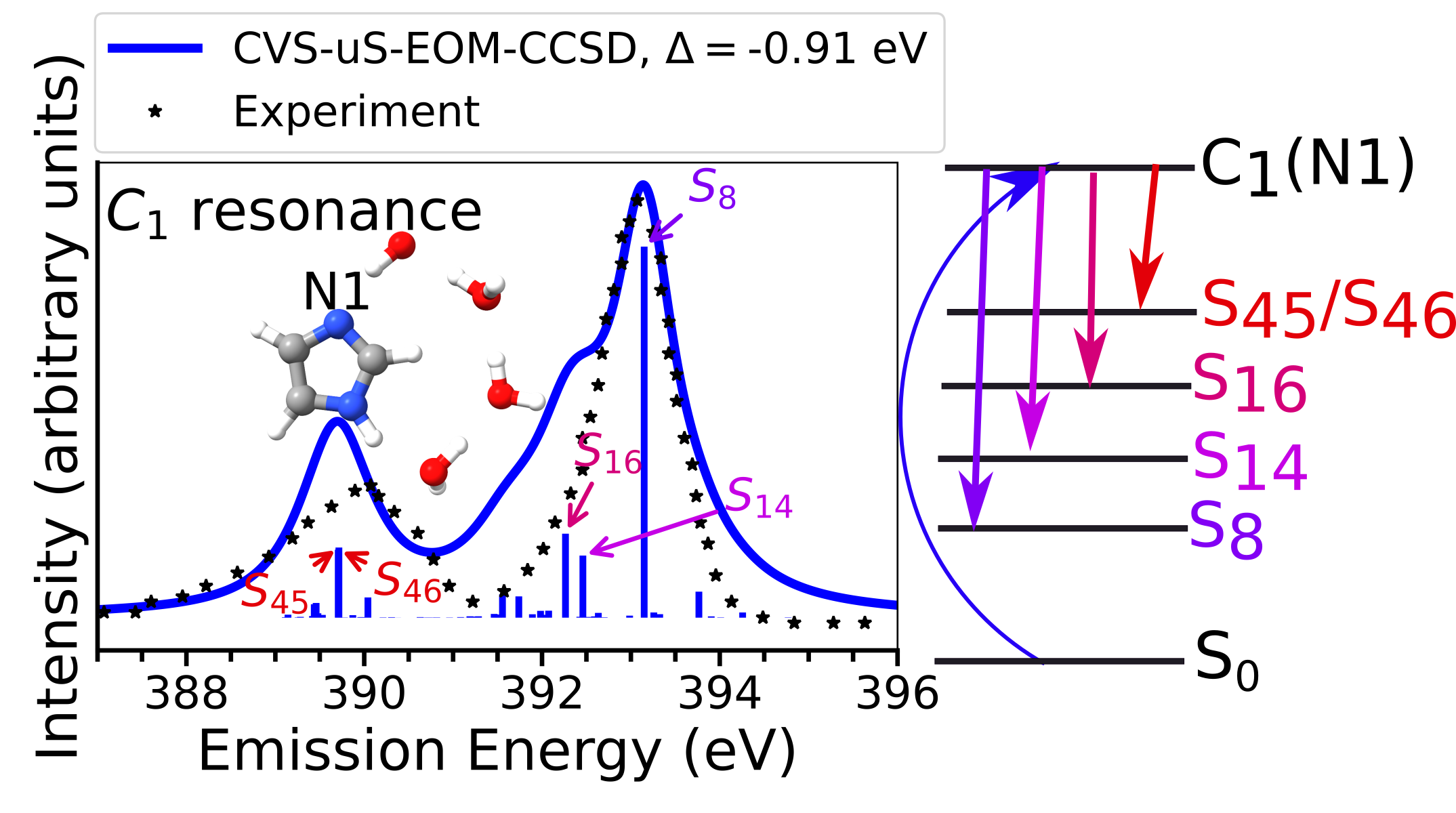}

\end{tocentry}
\nocite{}
\bibliography{bibliography}% Produces the bibliography via BibTeX.

\end{document}